\documentclass[conference]{IEEEtran}
\IEEEoverridecommandlockouts
\usepackage{cite}
\usepackage{amsmath,amssymb,amsfonts}
\usepackage{algorithmic}
\usepackage{graphicx}
\usepackage{textcomp}
\usepackage{xcolor}
\usepackage{listings}
\usepackage{tikz}
\usepackage[caption=false,font=footnotesize]{subfig} 
\usepackage[most]{tcolorbox}
\usetikzlibrary{patterns}
\definecolor{pastelgreen}{rgb}{0.47, 0.87, 0.47} 
\definecolor{babyblue}{rgb}{0.54, 0.81, 0.94} 
\definecolor{bananayellow}{rgb}{1.0, 0.88, 0.21}
\definecolor{babypink}{rgb}{0.96, 0.76, 0.76}

\definecolor{pastelred}{rgb}{1.0, 0.41, 0.38} 
\definecolor{lava}{rgb}{0.81, 0.06, 0.13}
\definecolor{brown}{rgb}{0.5, 0.16, 0.16}

\definecolor{blu1}{HTML}{90e0ef} 
\definecolor{blu2}{HTML}{0096c7}
\definecolor{blu3}{HTML}{023e8a}

\definecolor{red4}{HTML}{ffba08}
\definecolor{red3}{HTML}{f48c06}
\definecolor{red2}{HTML}{d00000}
\definecolor{red1}{HTML}{6a040f}

\definecolor{mlx4}{HTML}{ccff33}
\definecolor{mlx3}{HTML}{70e000}
\definecolor{mlx2}{HTML}{38b000}
\definecolor{mlx1}{HTML}{006400}

\newtcolorbox{observationbox}{
  colback=green!8,
  colframe=green!50!black,
  boxrule=0.8pt,
  arc=2mm,
  left=1.2mm,right=1.2mm,top=0.8mm,bottom=0.8mm
}

\def\BibTeX{{\rm B\kern-.05em{\sc i\kern-.025em b}\kern-.08em
    T\kern-.1667em\lower.7ex\hbox{E}\kern-.125emX}}
\begin{document}

\title{Characterizing the Impact of Congestion\\in Modern HPC Interconnects}

\author{
    \IEEEauthorblockN{
        Lorenzo Piarulli\IEEEauthorrefmark{1}, 
        Marco Faltelli\IEEEauthorrefmark{2}, 
        Dirk Pleiter\IEEEauthorrefmark{3}, 
        Karthee Sivalingam\IEEEauthorrefmark{3}\IEEEauthorrefmark{4}, 
        Dancheng Zhang\IEEEauthorrefmark{4}, \\
        Kexue Zhao\IEEEauthorrefmark{4}, 
        Matteo Turisini\IEEEauthorrefmark{5}, 
        Francesco Iannone\IEEEauthorrefmark{2}, 
        Aldo Artigiani\IEEEauthorrefmark{4}, 
        Daniele De Sensi\IEEEauthorrefmark{1}
    }
    \\[0.2cm]
    \IEEEauthorblockA{\IEEEauthorrefmark{1}Sapienza University of Rome, \{piarulli, desensi\}@di.uniroma1.it}
    \IEEEauthorblockA{\IEEEauthorrefmark{2}ENEA, \{marco.faltelli, francesco.iannone\}@enea.it}
    \IEEEauthorblockA{\IEEEauthorrefmark{3}OEHI, University of Groningen, dirk.pleiter@open-edge-hpc-initiative.org}
    \IEEEauthorblockA{\IEEEauthorrefmark{4}Huawei, \{karthee.sivalingam, zhangdancheng, zhaokexue, aldo.artigiani\}@huawei.com}
    \IEEEauthorblockA{\IEEEauthorrefmark{5}CINECA, m.turisini@cineca.it}
}

\maketitle

\begin{abstract}
High-performance computing (HPC) systems increasingly support both scalable AI training and large-scale simulation workloads. Both typically rely heavily on collective communication operations. On modern supercomputers, however, network congestion has emerged as a major limitation, driven by heterogeneous traffic patterns resulting from diverse workload mixes. As system scale and active users continue to grow, understanding how today’s interconnect technologies respond to congestion is essential for establishing realistic performance expectations and informing future system design. This paper presents a comprehensive characterization of congestion behavior across four major HPC fabrics: EDR InfiniBand, HDR InfiniBand, NDR InfiniBand, Cray Slingshot, and emerging Ethernet fabrics. These fabrics span high-performance proprietary interconnects as well as adaptive Ethernet-based designs aligned with emerging standards such as Ultra Ethernet. We evaluate their responses to both \textit{steady} congestion and a wide range of \textit{bursty} patterns that vary in duration, intensity, and pause length, capturing the \textit{bursty} communication typical of AI workloads. Our study covers multiple scales, examining how congestion manifests differently as system size increases and identifying scale-dependent behaviors that influence collective performance. By analyzing the challenges that arise under these controlled stress conditions, we aim to provide a practical overview of congestion issues and possible optimizations. The insights derived from this evaluation can guide researchers and HPC architects in designing more effective congestion-control mechanisms and network load-balancing strategies.
\end{abstract}


\section{Introduction}
Supercomputers are rapidly increasing in both scale and performance, enabling new advances in scientific computing and artificial intelligence. As system sizes grow and user demand increases, multiple jobs are often executed concurrently, with workloads running either on dedicated or shared compute nodes. Moreover, with the growth of the number of endpoints, new network topologies started to become popular that can be expected to be more susceptible to congestion. Regardless of the allocation strategy, all applications ultimately rely on a shared communication infrastructure. When several jobs simultaneously traverse common network components, links may become oversubscribed and switch buffers can fill, leading to network congestion. Since communication is a critical phase of both HPC and AI workloads, congestion directly translates into performance degradation and reduced system efficiency. Although congestion can affect any HPC system, its manifestation and impact depend on several factors, including network topology, system scale, traffic patterns, communication primitives, fabric technology, and the congestion-mitigation mechanisms available. To address these challenges, a wide range of solutions has been proposed. These include end-host–based and network-level congestion control mechanisms that detect congestion signals and throttle injection rates, as well as other network-level techniques, such as adaptive routing and load balancing, that mitigate congestion.

The problem is further complicated by dynamic, multi-tenant workloads and allocation policies optimized for overall system utilization. As a result, network resources are frequently shared, and congestion must be detected and mitigated at runtime. Modern congestion-control and load-balancing mechanisms attempt to address this challenge by dynamically reallocating flows, adjusting routing decisions, or reducing traffic injection rates at the sources, thereby improving traffic distribution and limiting packet loss.

Therefore, the ability to effectively mitigate congestion is tightly coupled with the interconnect technology deployed by an HPC facility. The more capable a technology is at handling diverse congestion scenarios, the better the system can support concurrent multi-tenant workloads with heterogeneous communication patterns. However, modern supercomputing environments rely on a wide variety of interconnects, each with distinct architectural designs, routing strategies, and congestion-management mechanisms.

In recent years, alongside the strong performance and widespread adoption of NVIDIA’s InfiniBand, several Ethernet-based solutions have gained increasing attention. This trend has been driven both by the emergence of the Ultra Ethernet Consortium\cite{ultra} and by the demonstrated effectiveness of Ethernet-based technologies such as HPE/Cray's Slingshot, which now account for a significant fraction of the aggregate performance of systems listed in the TOP500 ranking \cite{TOP500}.
Given the diversity of interconnect technologies and congestion-management solutions deployed in modern supercomputing systems, it is essential to systematically examine their limitations, characterize congestion effects, and correlate these effects with specific architectural and protocol-level features. A comprehensive and comparative analysis is therefore required to understand how different fabrics behave under realistic congestion conditions.

In this work, we analyze multiple state-of-the-art interconnect technologies, including EDR, HDR, and NDR InfiniBand, Cray Slingshot, and \textit{Network Scale Load Balance} (NSLB) enabled Ethernet fabrics. We evaluate these technologies under a range of controlled congestion scenarios generated using collective communication patterns with both persistent and \textit{bursty} traffic intensities, designed to emulate realistic HPC and AI workloads.
Our evaluation spans both production-scale supercomputers and research testbeds. Specifically, we analyze CINECA’s Leonardo, ENEA’s CRESCO8, and LUMI systems at both small and large scales, and we additionally evaluate emerging Network
Scale Load Balance NSLB-capable Ethernet fabrics deployed in the HAICGU and Nanjing laboratory clusters.
Through this evaluation, we characterize the behavior of modern supercomputing networks under congestion, identify their distinguishing features, and assess the effectiveness of their congestion-mitigation mechanisms.

\section{Background}\label{sec:back}

Congestion manifests in distinct forms, each requiring a specific mitigation strategy. As messages traverse multiple switches between source and destination, the physical location of the bottleneck dictates the appropriate response. When congestion develops at an \textit{intermediate} switch, \textbf{load balancing} (or adaptive routing) can effectively alleviate the pressure by rerouting traffic along alternative paths to avoid resource collisions. Conversely, if congestion occurs at an \textit{edge} switch directly connected to the endpoint, {load balancing} becomes ineffective because all possible network paths must eventually converge on that specific switch to reach the destination. In such scenarios, \textbf{congestion control} mechanisms are essential to throttle the injection rate at the source. 

These dynamics are closely tied to specific communication patterns: \emph{\emph{AlltoAll}} and \textit{permutation} communications frequently stress \textit{intermediate} switches due to global bandwidth demands, whereas \emph{Incast} patterns, where multiple nodes simultaneously target a single receiver, typically lead to congestion at the \textit{edge}.
In our analysis, we consider five different systems (described in Table \ref{tab:env_summary}) based on the three main existing interconnect technologies: Ethernet, InfiniBand, and Slingshot, which we describe in the following.

\subsection{Ethernet}
Ethernet’s popularity in large-scale clusters stems from its open, standards-based ecosystem, which fosters vendor competition, ensures broad \textit{interoperability}, and significantly reduces the total cost of ownership. By avoiding vendor lock-in and leveraging mature management tools, organizations can scale infrastructure while closing the performance gap with specialized fabrics through technologies like \textit{RDMA over Converged Ethernet} (RoCE) and the \textit{Ultra Ethernet Consortium} (UEC) \cite{ultra}. Specifically, the \textit{Ultra Ethernet Specification v1.0} \cite{uec_spec} introduces the \textit{Ultra Ethernet Transport} (UET), which brings HPC-grade capabilities, such as multipath packet spraying, out-of-order delivery, and local link-layer retransmissions, to the standard IP ecosystem to mitigate tail latency.

\textbf{Congestion Control:} While traditionally a "best-effort" medium, Ethernet has evolved to support the \textit{lossless} operation required by RDMA through \textit{Priority Flow Control} (PFC), which prevents packet loss by pausing specific traffic classes at the link level, though it can trigger \textit{deadlock} or \textit{congestion spreading}~\cite{zhu2015dcqcn}. To mitigate these risks, modern fabrics employ proactive \textit{congestion control} typically based on \textit{Explicit Congestion Notification} (ECN), where switches mark IP headers upon exceeding queue thresholds so that the source can throttle its injection rate before buffers overflow. Building on this, the congestion control algorithm \textit{Data Center Quantized Congestion Notification} (DCQCN) couples ECN marking with \textit{Quantized Congestion Notification} (QCN) style feedback to rapidly restore fairness \cite{zhu2015dcqcn}, while \textit{TIMELY} utilizes fine-grained RTT gradients to bound queuing delay without switch-side marking \cite{mittal2015timely}. More advanced schemes like \textit{HPCC} leverage in-network telemetry to compute near-optimal rate updates, achieving fast convergence under large \emph{Incast} \cite{li2019hpcc}. Within the \textit{Ultra Ethernet} framework, congestion control leverages both receiver- and sender-based algorithms \cite{bonato2024smartt-anon}, and the Huawei \textit{Cloud Engine} switches (CE8850 and CE9855) analyzed in this work implement \textit{AI ECN} to dynamically adjust ECN thresholds based on the workload~\cite{huaweiAIECN}.

\textbf{Load Balancing:} Ethernet routing strategies range from \textit{congestion-oblivious} methods like \textit{Equal-Cost Multi-Path} (ECMP) \cite{ecmp}, which frequently suffers from hash collisions and link underutilization \cite{hedera,conga,facebook,ecnsu}, to \textit{congestion-aware} dynamic adaptation. While schemes such as \textit{MPTCP} \cite{mptcp}, \textit{PLB} \cite{plb}, \textit{FlowBender} \cite{flowbender}, \textit{Flowlet Switching} \cite{flowlet}, \textit{Flowcell} \cite{flowcell}, and \textit{Flowcut} \cite{flowcut} improve efficiency by splitting traffic into subflows or flowlets, they often introduce state overhead and potential packet reordering \cite{Lu2018}. More granular packet-level approaches like \textit{OPS} \cite{Dixit2013} and \textit{MPRDMA} \cite{Lu2018} maximize path utilization through spraying but require robust out-of-order delivery handling. To address these challenges, the \textit{Huawei Cloud Engine} switches examined in this work utilize \emph{Network Scale Load Balance} (NSLB), a decentralized, real-time mechanism integrated into the network controller \cite{Wang2025_NetworkLoadBalancing}. By employing a \emph{flow matrix} to compute collision-free uplink assignments for each $(\textit{source edge}, \textit{destination edge})$ pair, NSLB ensures that concurrent flows from the same source are distributed across distinct uplinks, effectively minimizing collisions and optimizing fabric throughput \cite{Wang2025_NetworkLoadBalancing}.


\subsection{InfiniBand}
InfiniBand is a tightly integrated interconnect fabric that has evolved over multiple generations to provide the high bandwidth and low latency required for traditional HPC and AI workloads. While its specialized architecture offers superior performance metrics, it can lead to vendor lock-in and lacks the broad interoperability found in commodity Ethernet ecosystems.

\textbf{Congestion Control:} Unlike Ethernet's reactive pause-based mechanisms, InfiniBand employs a hardware-level, hop-by-hop credit-based flow control that inherently avoids buffer overflows by ensuring a sender only transmits when the receiver has available buffer space. However, this can lead to backpressure that propagates through the fabric, potentially creating "victim" flows. To mitigate this, InfiniBand utilizes an end-to-end closed-loop mechanism where congested switch ports probabilistically mark packets with \textit{Forward Explicit Congestion Notification} (FECN). The destination then returns a \textit{Backward Explicit Congestion Notification} (BECN), or \textit{Congestion Notification Packets} (CNPs), to the source, which throttles its injection rate via per-flow inter-packet delay. This makes the fabric's performance highly sensitive to the precise tuning of marking thresholds and recovery rates \cite{gran2011ibcc,alali2017ibcongestion}.

\textbf{Load Balancing:} While InfiniBand was historically built on deterministic forwarding, modern implementations provide \emph{Adaptive Routing} (AR) to dynamically manage link utilization. In an AR-enabled fabric, switches independently select among groups of equivalent egress ports based on real-time port load. This dynamic selection is orchestrated by the \textit{Subnet Manager} (SM), which is responsible for configuring AR groups and implementing topology-specific routing policies, such as those optimized for fat-trees or Dragonfly+, to maximize bisection bandwidth and minimize contention \cite{RocherGonzalezCCGrid22}.

\subsection{Slingshot}
HPE/Cray Slingshot follows a distinct design philosophy by merging the high-performance characteristics of specialized HPC fabrics like InfiniBand with the openness and interoperability of the Ethernet ecosystem. By optimizing \textit{RDMA over Ethernet} (RoCE), Slingshot achieves performance levels comparable to proprietary interconnects while remaining compatible with standard data-center infrastructure. This approach has catalyzed a broader industry trend where vendors seek to leverage existing Ethernet-based management tools and expertise without sacrificing the low-latency requirements of large-scale AI and HPC clusters.

\textbf{Congestion Control:} Slingshot implements a hardware-based \textit{congestion management} system designed to provide granular flow isolation and prevent the "congestion spreading" typical of traditional lossless Ethernet. Unlike \textit{Priority Flow Control} (PFC), which operates on broad traffic classes and can lead to head-of-line blocking, Slingshot tracks every packet and provides feedback for thousands of individual flows. This allows the fabric to identify and throttle only the specific sources contributing to a bottleneck, ensuring that "victim" flows remain unaffected. By maintaining low, stable queuing delays even under heavy \emph{Incast} conditions, Slingshot provides the performance predictability required for tightly coupled parallel workloads \cite{slingshot,RowethCUG22Slingshot}.

\textbf{Load Balancing:} Slingshot couples fine-grained \textit{adaptive routing} with global congestion awareness to optimize path selection, particularly in Dragonfly deployments. Switches dynamically evaluate multiple minimal and non-minimal paths by estimating real-time path congestion and considering path length, allowing for high bisection bandwidth utilization. This mechanism is especially effective in multi-tenant environments, as it can steer traffic away from transient hotspots and minimize tail-latency interference caused by competing workloads. Furthermore, Slingshot can guarantee in-order delivery of packets even in the presence of adaptive routing by draining active flows before changing their path \cite{flowcut}.

\section{Methodology}

We developed a custom experimental pipeline that injects controlled congestion while executing collective communication operations and simultaneously benchmarking their performance. The standard way to create congestion is to split the allocated nodes into \emph{victims} and \emph{aggressors}, and then measure the victims while the aggressors generate continuous, communication-intensive traffic \cite{slingshot, GPCNeT}. We started from this idea, but then developed a more flexible approach capable of producing multiple congestion scenarios. In addition to steady, continuous congestion, we also inject \emph{bursty} congestion, varying both the burst duration and the idle gap between bursts to emulate realistic, time-varying contention. By implementing various congestion profiles, we can simulate a spectrum of network conditions and analyze the system's response to each scenario.

While the aggressors are running, the victims perform 1,000 iterations of the benchmarked collectives, all of which are recorded. We then discard the first 100 iterations to account for network warmup. The final execution time used to compare congested collectives against uncongested ones is computed by taking the mean of the remaining iterations.

The following chapter describes the methods used to carry out this analysis in detail.

\subsection{Congestion Injection}

In our approach, the aggressors continuously launch collectives without interruption in an endless loop, thereby generating communication noise. The victims, instead, perform the benchmarked collectives for a fixed number of iterations. Group creation is performed by interleaving the nodes in increasing order: the first node is assigned to the victims’ group, the second to the aggressors’ group, the third again to the victims’ group, and so on, alternating until all nodes are allocated. This ensures a balanced distribution of aggressors and victims across the allocated set of nodes, maximizing network resources sharing and, thus, congestion. The benchmark is first executed on the victims without congestion to
establish a baseline. Then, a congestion job and the benchmark are run in parallel on the aggressors and victims, respectively, to measure and analyze the effects of congestion. Attackers are provided with two types of collectives for noise injection: \emph{AlltoAll} and \emph{Incast}. The first is used to send as many messages as possible to all nodes, creating a general state of network noise. The latter, instead, focuses the traffic on a
single node, attempting to generate congestion at a specific \textit{edge} switch.

\subsection{Selected Collectives}

\begin{figure}
\centering
\begin{tikzpicture}
    \usetikzlibrary{patterns} 

    \def\Total{6}
    \def\Height{0.37}
    \def\Spacing{1.2} 
    
    \def\Aa{0.3401}
    \def\Ca{0.6599}
    
    \def\Ab{0.2230}
    \def\Bb{0.3281}
    \def\Cb{0.4489}
    
    \fill[lava, postaction={pattern=dots, pattern color=black}] 
        (0,0) rectangle (\Aa*\Total,\Height);
    \fill[pastelgreen, postaction={pattern=vertical lines, pattern color=black}] 
        ({(\Aa)*\Total},0) rectangle (\Total,\Height);
    \draw[thick] (0,0) rectangle (\Total,\Height);
    
    \node at (\Aa*\Total/2,\Height+0.2) {Memory};
    \node at (\Aa*\Total+\Ca*\Total/2,\Height+0.2) {Communication};
    
    \node[left] at (-0.2,\Height/2) {\emph{AlltoAll}};
    
    \fill[lava, postaction={pattern=dots, pattern color=black}] 
        (0,-\Spacing) rectangle (\Ab*\Total,-\Spacing+\Height);
    \fill[bananayellow, postaction={pattern=horizontal lines, pattern color=black}] 
        (\Ab*\Total,-\Spacing) rectangle ({(\Ab+\Bb)*\Total},-\Spacing+\Height);CI TEiMAster NCE
    \fill[pastelgreen, postaction={pattern=vertical lines, pattern color=black}] 
        ({(\Ab+\Bb)*\Total},-\Spacing) rectangle (\Total,-\Spacing+\Height);
    \draw[thick] (0,-\Spacing) rectangle (\Total,-\Spacing+\Height);
    
    \node at (\Ab*\Total/2,-\Spacing+\Height+0.2) {Memory};
    \node at (\Ab*\Total+\Bb*\Total/2,-\Spacing+\Height+0.2) {Reduction};
    \node at (\Ab*\Total+\Bb*\Total+\Cb*\Total/2,-\Spacing+\Height+0.2) {Communication};
    
    \node[left] at (-0.2,-\Spacing+\Height/2) {\emph{AllReduce}};

\end{tikzpicture}
\caption{Comparison of time distribution between \emph{AlltoAll} and \emph{AllReduce} operations.}
\label{fig:memory_handling}
\end{figure}
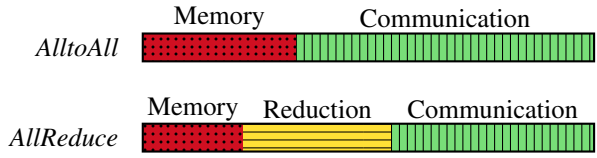

Our analysis is based on collective communication primitives, which are the \textit{de facto} methodology for distributed computing in HPC systems. These collectives are typically provided by MPI libraries such as NCCL, RCCL, Open MPI and MPICH, based on the system library availability. A wide range of collectives exists, each serving a different purpose: some are purely communication-oriented (e.g., \emph{AllGather}, \emph{AlltoAll}, and Broadcast), while others also include computation through reduction operations (e.g., \emph{AllReduce} and \emph{ReduceScatter})
Our initial experiments considered the Open MPI \cite{OpenMPI} \emph{AllReduce} collective, which showed up to a 25\% bandwidth loss compared to \emph{AlltoAll} \cite{10.1109/SCW63240.2024.0012}. A follow-up analysis with a custom ring \emph{AllReduce} (separating \emph{ReduceScatter} and \emph{AllGather}) indicated that performance was mainly limited by reduction costs and memory handling (initial buffer setup and memcpy operations), rather than by network communication (Figure \ref{fig:memory_handling}). Since these overheads also affect other collectives and can dominate execution time, we exclude computation-based collectives (\emph{AllReduce} and \emph{ReduceScatter}) from the rest of this work. We therefore focus on communication-only collectives (\emph{AlltoAll} and \emph{AllGather}) and measure only their communication phase, enabling a cleaner characterization of network behavior and congestion effects. To eliminate the uncertainty introduced by the MPI libraries' dynamic algorithm selection and ensure a fair comparison across different software stacks, we did not use the default MPI collective implementations. We implemented custom ring \emph{AllGather} and linear \emph{AlltoAll} algorithms via standard MPI \emph{send}/\emph{recv} primitives. This approach ensured we used the same communication pattern across all systems, since they rely on different MPI libraries (Cray MPI on LUMI, Open MPI on Leonardo/CRESCO8). Moreover, our custom implementation allowed us to remove memory-handling overheads such as malloc and memcpy of temporary buffers, ensuring that measurements focus strictly on network-level communication latency and reducing the impact of differences in node and memory architectures.

\subsection{Steady Congestion Injection}

The first method injects \emph{steady}, continuous congestion using the \emph{AlltoAll} and \emph{Incast} collectives. By repeatedly executing the same collective pattern over an extended time window, we create a quasi-stationary stress condition in which queue occupancy, link utilization, and contention hotspots remain stable (or oscillate within a narrow range) for most of the run. This “steady-state” regime is useful because it exposes the behavior of routing, arbitration, and flow-control mechanisms once initial transients have decayed. In particular, persistent congestion gives time for closed-loop mechanisms (e.g., ECN/credit-based backpressure and rate adaptation) to converge, allowing us to observe whether the fabric reaches a stable operating point (bounded queues and fair throughput) or instead exhibits pathological effects such as sustained head-of-line blocking, victim-flow throughput collapse, or congestion spreading.

The \emph{AlltoAll} collective primarily exercises \emph{transit contention} by simultaneously activating many source--destination pairs whose shortest paths overlap on \textit{intermediate} links and core/aggregation switches. When multiple minimal routes exist, this condition is well suited to evaluate \emph{load balancing and adaptive routing}, because congestion can often be mitigated by shifting traffic to alternative paths with lower instantaneous occupancy. Conversely, \emph{Incast} concentrates traffic from many senders onto a single receiver (or a small set of receivers), creating a \emph{fan-in bottleneck} at the destination NIC and the adjacent edge (ToR) switch. Since the limiting resource is typically the receiver-side egress and buffering rather than a congested transit link, rerouting cannot eliminate the hotspot. As a result, \emph{Incast} congestion isolates the effectiveness of \emph{end-host and switch-level congestion control} (rate reduction, marking thresholds, and recovery dynamics), and highlights whether the fabric can prevent queue build-up and backpressure from propagating to unrelated flows.

\subsection{Bursty Congestion Injection}\label{sec:cong:bursty}

The second method generates congestion in \emph{bursts} rather than continuously. Each burst consists of one or more consecutive collective invocations, and we vary both the burst duration (i.e., number of collectives per burst or total burst time) and the \emph{inter-burst idle interval}. These parameters control how far congestion can develop within device buffers and switch queues before traffic stops, and how much time the network has to drain queues and restore \textit{steady} throughput before the next burst. By sweeping multiple burst/idle combinations, we capture regimes ranging from “microbursts” that may only transiently fill shallow buffers to longer bursts that can trigger feedback-based throttling and potentially affect subsequent communication phases.

Bursty congestion is inherently \emph{intermittent} and therefore stresses detection and reaction latency: if congestion onset is faster than the control loop (marking, feedback, and sender rate adaptation), the system may only respond after the burst has already completed, leaving residual queueing delay and backpressure that impacts other traffic. Moreover, each burst effectively reintroduces a transient, forcing the network to repeatedly transition between uncongested and congested states; this exposes overshoot/undershoot behavior, recovery speed, and stability under repeated excitation. This pattern closely resembles communication phases in production HPC and AI workloads, such as distributed deep learning, where gradient aggregation (e.g., \emph{AllReduce}) follows each optimization step and produces periodic spikes in network utilization rather than a constant load.

\begin{figure}[htbp]
    \centering
    \includegraphics[width=1\linewidth]{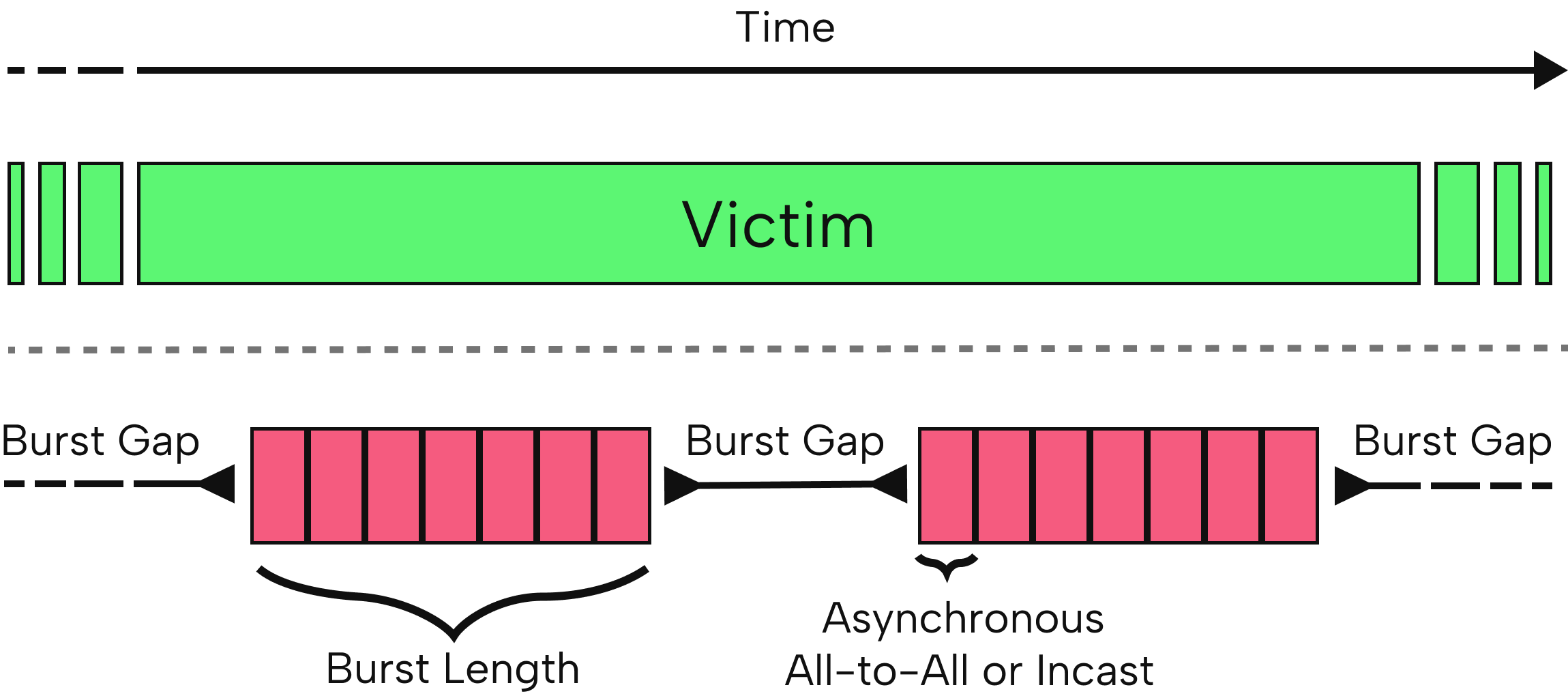}
    \caption{\textit{Bursty} congestion injection visualization, \textit{bursty} aggressor on the bottom and victim on the top.}
    \label{fig:haicgu_scatter}
\end{figure}

\subsection{Evaluation Environments}

The fabrics have been evaluated under many different architectures, node counts, and topologies. This allowed us to understand congestion under a variety of scenarios. The systems considered were CINECA's Leonardo, ENEA's CRESCO8, LUMI, Huawei AI and Computing at Goethe University (HAICGU), and the HPC lab in Nanjing (Nanjing).
Leonardo, CRESCO8, and LUMI are supercomputers ranked in the TOP500 \cite{TOP500}, offering large-scale executions and production environments with many active users. On these systems, we cannot fully control job allocations, as they depend on the current usage of the computers. These systems enforce different policies on the maximum allocation size; therefore, we evaluated them at multiple scales during testing. For comparison and continuity, in this paper we report the 256 node results as maximum allocations, since it is a common scale across the platforms and the maximum available on Leonardo.

HAICGU and the Nanjing lab, in contrast, are research HPC systems designed for evaluating new architectures. They are small, with a maximum of 10 nodes for HAICGU and 8 nodes for Nanjing. Despite their small size, they provide the ability to test the emerging Ethernet fabrics CE8850 and CE9855, and to compare it with an EDR InfiniBand system composed of the same TaiShan nodes.

\vspace{1em}

\textbf{Leonardo}
is composed of two partitions: GPU-accelerated (Booster) and CPU-only partition (Data Centric General Purpose)~\cite{turisini2023leonardopaneuropeanpreexascalesupercomputer}. We use the Booster partition which employs BullSequana X2135 “Da Vinci” nodes featuring a single Intel Xeon Platinum 8358 CPU (32 cores, 2.60 GHz) paired with 4$\times$ NVIDIA A100 GPUs (64 GB HBM2e) and 512 GB RAM(8$\times$64 GB DDR4) per node. All compute nodes are interconnected via NVIDIA Mellanox HDR InfiniBand with a Dragonfly+ topology. Every node includes 2 dual port HDR100 boards delivering 400Gb/s per node.

\vspace{1em}

\textbf{CRESCO8}
\cite{cresco} is the new Tier-0 HPC facility at ENEA. This cluster comprises 760 CPU nodes and 17 GPU nodes. For the massive scale purposes of this paper, we chose to run our experiments on the CPU partition. Each CPU node is composed of 2$\times$64 Intel Xeon Platinum 8592+ cores, 512 GB RAM and a dual-port Mellanox-NDR ConnectX-7 NIC that delivers 200 Gb/s. Nodes are connected through a 1.67:1 blocking Fat-Tree topology.

\vspace{1em}

\textbf{LUMI}
uses HPE Cray EX nodes in a multi-partition architecture based on AMD CPUs and GPUs. We selected the GPU partition, which consists of 2978 GPU-CPU nodes, each equipped with one AMD EPYC “Trento” CPU, four AMD Instinct MI250X accelerators, and 512 GB of system DDR4 memory. The accelerators provide a total of 512 GB of HBM2e memory per node. All nodes are interconnected through the Cray Slingshot fabric using 4$\times$200 Gb/s links in a Dragonfly topology, enabling scalable communication across the system\cite{LUMI_website}.

\vspace{1em}

\textbf{HAICGU}
uses TaiShan 200 (Model 2280) nodes with dual Kunpeng 920 CPUs (ARMv8 AArch64, 64 cores, 2.6GHz) and 128GB RAM (16$\times$8GB DIMMs).
It is maintained by Open Edge and HPC initiative (OEHI) \cite{OEHI_website}, has two 10-node partitions with 2 leaf switches each: one using Mellanox MSB7890-ES2F EDR switches and the other RoCE-based CE8850, both at 100GE.

\vspace{1em}

\textbf{Nanjing}
uses TaiShan 200 (Model 2280) nodes with dual Kunpeng 920 CPUs (ARMv8 AArch64, 64 cores, 2.6GHz) and 128GB RAM (16$\times$8GB DIMMs).
It includes 8 nodes on a RoCE partition built with 4 CE9855 switches \textbf{with NSLB}, arranged in a two-leaf, two-spine 200GE topology. This design allows NSLB to leverage multiple path configurations for load balancing.

\begin{table*}[t]
\centering
\caption{Summary of evaluated environments and interconnect configurations.}
\label{tab:env_summary}
\renewcommand{\arraystretch}{1.15}
\setlength{\tabcolsep}{4.5pt}
\begin{tabular}{p{1.8cm} p{1.3cm} p{2.0cm} p{4.2cm} p{1.4cm} p{2.6cm} p{3.1cm}}
\hline
\textbf{System} & \textbf{Partition Nodes} & \textbf{Interconnect} & \textbf{Compute node (CPU/GPU)} & \textbf{Memory} & \textbf{Link rate / node} & \textbf{Topology} \\
\hline
Leonardo (CINECA) & 3456 & HDR InfiniBand & BullSequana X2135 ``Da Vinci'', Intel Xeon Platinum 8358 & 512\,GB & 400\,Gb/s (2$\times$dual-port) & Dragonfly+ \\
\hline
CRESCO8 (ENEA) & 760 & NDR InfiniBand & Intel Xeon Platinum 8592+ & 512\,GB & 200\,Gb/s (dual-port) & 1.67:1 blocking Fat-Tree \\
\hline
LUMI (CSC) & 2978 & Cray Slingshot & AMD EPYC ``Trento'' & 512\,GB & 800\,Gb/s (4$\times$200\,Gb/s links) & Dragonfly\\
\hline
HAICGU & 10 & EDR InfiniBand/RoCE & TaiShan 200 (2280), Kunpeng 920 & 128\,GB & 100\,GE; & Single switch \\
\hline
Nanjing lab & 8 & RoCE-NSLB & TaiShan 200 (2280), Kunpeng 920 & 128\,GB  & 200\,GE &  2-spine/2-leaf\\
\hline
\end{tabular}
\end{table*}

\subsection{Analysis Structure}
To build the results section progressively, we start from the least challenging scenario and move toward the most demanding one. Specifically, we first analyze \textit{steady} congestion at small scale (Sec.\ref{sec:results:small}), where contention is limited and congestion effects are typically easier to mitigate. Although small-scale setups are not the most common setting in which severe congestion emerges, they are still useful to (i) characterize baseline congestion behavior in controlled scenarios and (ii) provide an initial, clean evaluation of emerging Ethernet-based fabrics before turning to large-scale (Sec.\ref{sec:results:large}) and \textit{bursty} (Sec.\ref{sec:results:bursty}) regimes, which constitute the main focus of this paper.

\section{Steady Congestion at Small Scale}\label{sec:results:small}

\subsection{HAICGU and Nanjing}
Since prior work reported a throughput drop for the CloudEngine (CE) switches under \emph{\emph{AlltoAll}} communication \cite{10.1109/SCW63240.2024.0012}, we first compare the two CE-based Ethernet testbeds under uncongested conditions to further investigate the reasons for this drop. On HAICGU (CE8850), RoCE cannot sustain throughput with large messages due to a recurring \textit{sawtooth} pattern for \emph{AlltoAll} and \emph{AllGather} on vectors larger than 16~MiB (Figure~\ref{fig:haicgu_scatter}). On the same system (thus using the same compute nodes) InfiniBand remains stable, and we can thus attribute such behavior to an unstable feedback-based congestion-control response \cite{CHIU19891}. In contrast, the Nanjing CE9855-based fabric maintains stable throughput across message sizes and does not exhibit the oscillations seen on the CE8850 (Figure~\ref{fig:haicgu_scatter}), achieving variability comparable to that of the HAICGU InfiniBand partition. This indicates that the issue has been resolved in later generations of CE switches. For these reasons, we exclude the HAICGU system, based on CE8850 switch, from the following analysis due to its unstable behavior.

\begin{figure}[htbp]
    \centering
    \includegraphics[width=1\linewidth]{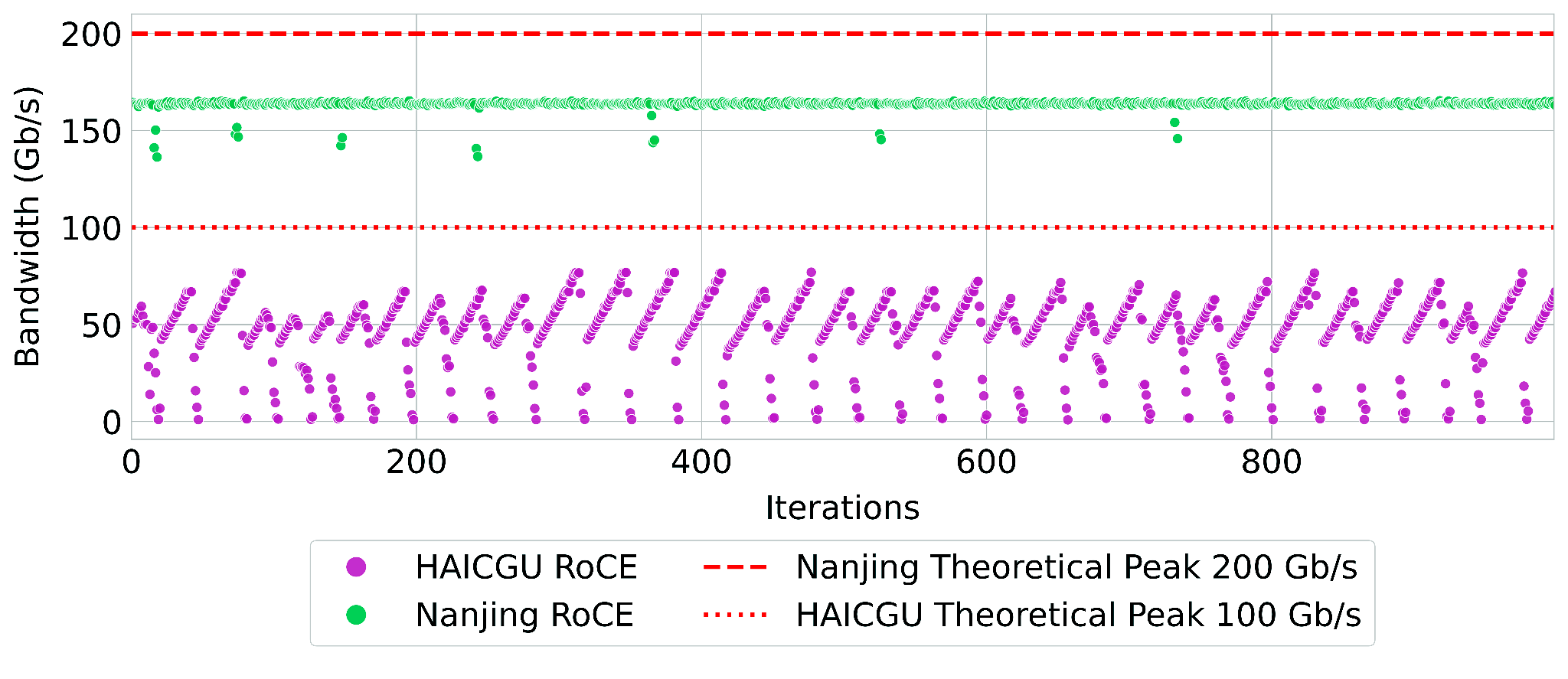}
    \caption{4 nodes HAICGU sawtooth behavior on 128 MiB messages with \emph{AllGather} collective}
    \label{fig:haicgu_scatter}
\end{figure}

\begin{observationbox}
\textbf{Observation 1.} Even without explicit congestion injection, a system can still exhibit congestion-related effects: if an application’s communication rate is sufficiently high, the congestion control may be unable to sustain it, leading to throughput drops and instability even when no other workloads are present.
\end{observationbox}

\begin{figure}[htbp]
    \centering
    \includegraphics[width=1\linewidth]{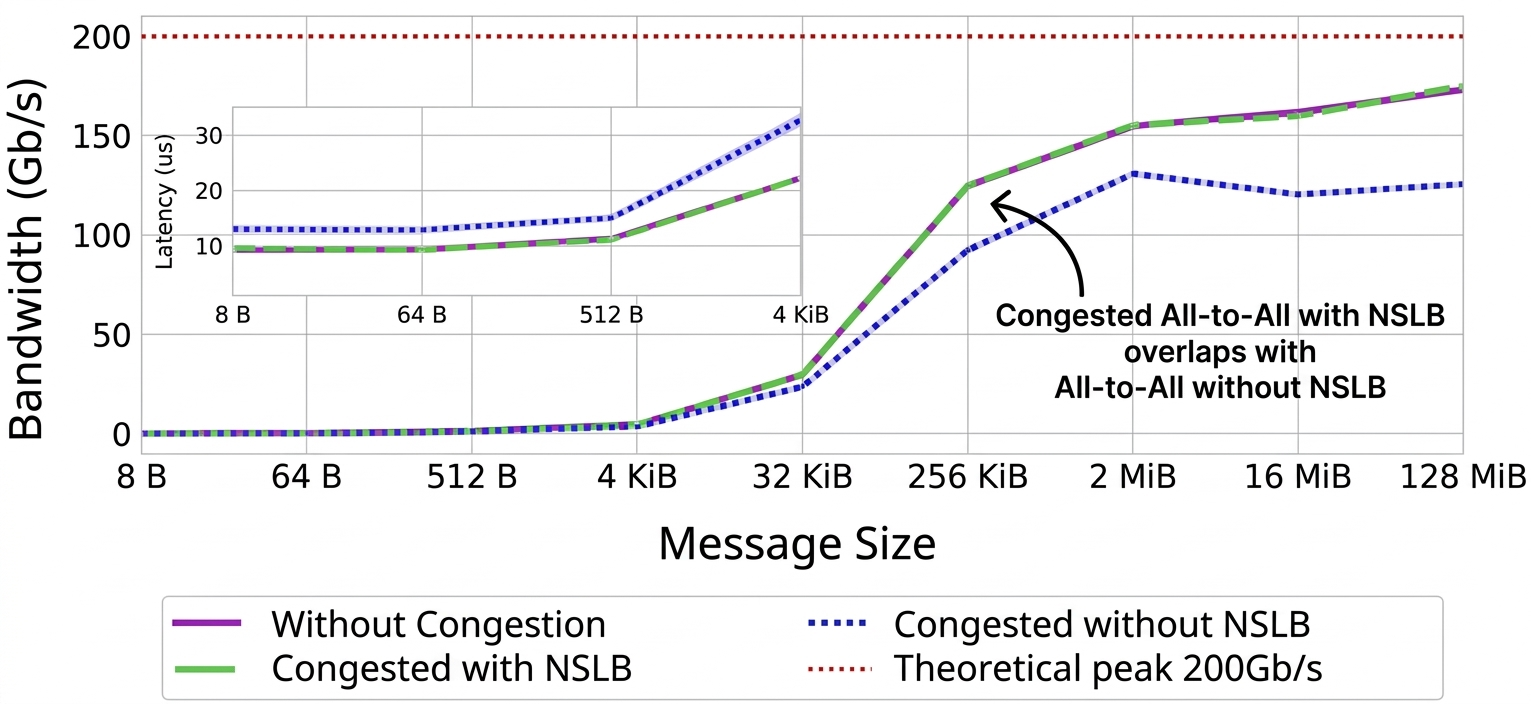}
    \caption{\textit{Steady} NSLB analysis in a \emph{AlltoAll} congestion with 4 victims and 4 aggressor nodes}
    \label{fig:plot_nslb}
\end{figure}

On the Nanjing system, based on the CE9855 Ethernet switch, we instead also analyze the impact of congestion on network performance. On this testbed, we can also explicitly enable and disable NSLB to isolate the contribution of network load balancing, and see how the system would behave without load balancing. We run an \emph{AlltoAll} as victim workload on four nodes, and another \emph{AlltoAll} as aggressor on the other four nodes. Nodes are allocated to the victim and the aggressor in an interleaved way. Figure~\ref{fig:plot_nslb} shows the results of the analysis. Without congestion, the victim \emph{AlltoAll} reaches a peak bandwidth of 180~Gb/s. When NSLB is active, there is no performance drop even in presence of congestion (the two lines perfectly overlap). On the other hand, when disabling NSLB, in presence of congestion the performance drops to 120~Gb/s, showing the effectiveness of NSLB.

\subsection{Leonardo, CRESCO8, and LUMI}
At the same scale, Leonardo, CRESCO8, and LUMI exhibit standard uncongested behavior, without signaling particular drops or anomalies. 
Even in presence of congestion, they handle \textit{steady} congestion with limited performance impact. For Leonardo and LUMI, the Dragonfly+/Dragonfly topology is particularly effective for small allocations, since nodes are often distributed across groups, enabling multiple paths between endpoints; as a result, load balancing can easily deflect traffic and resolve congestion on \textit{intermediate} switches. Despite its different topology, CRESCO8 exhibits similarly robust behavior, and its blocking topology does not noticeably affect performance at this scale. Overall, these results indicate that, when operating at small scale on sufficiently large systems, \textit{steady} congestion is often mitigated by path diversity and available network capacity, reducing the sensitivity to the specific topology or congestion-control mechanism.

\begin{figure*}[!t]
  \centering
  \subfloat[CRESCO8]{\includegraphics[width=0.32\textwidth]{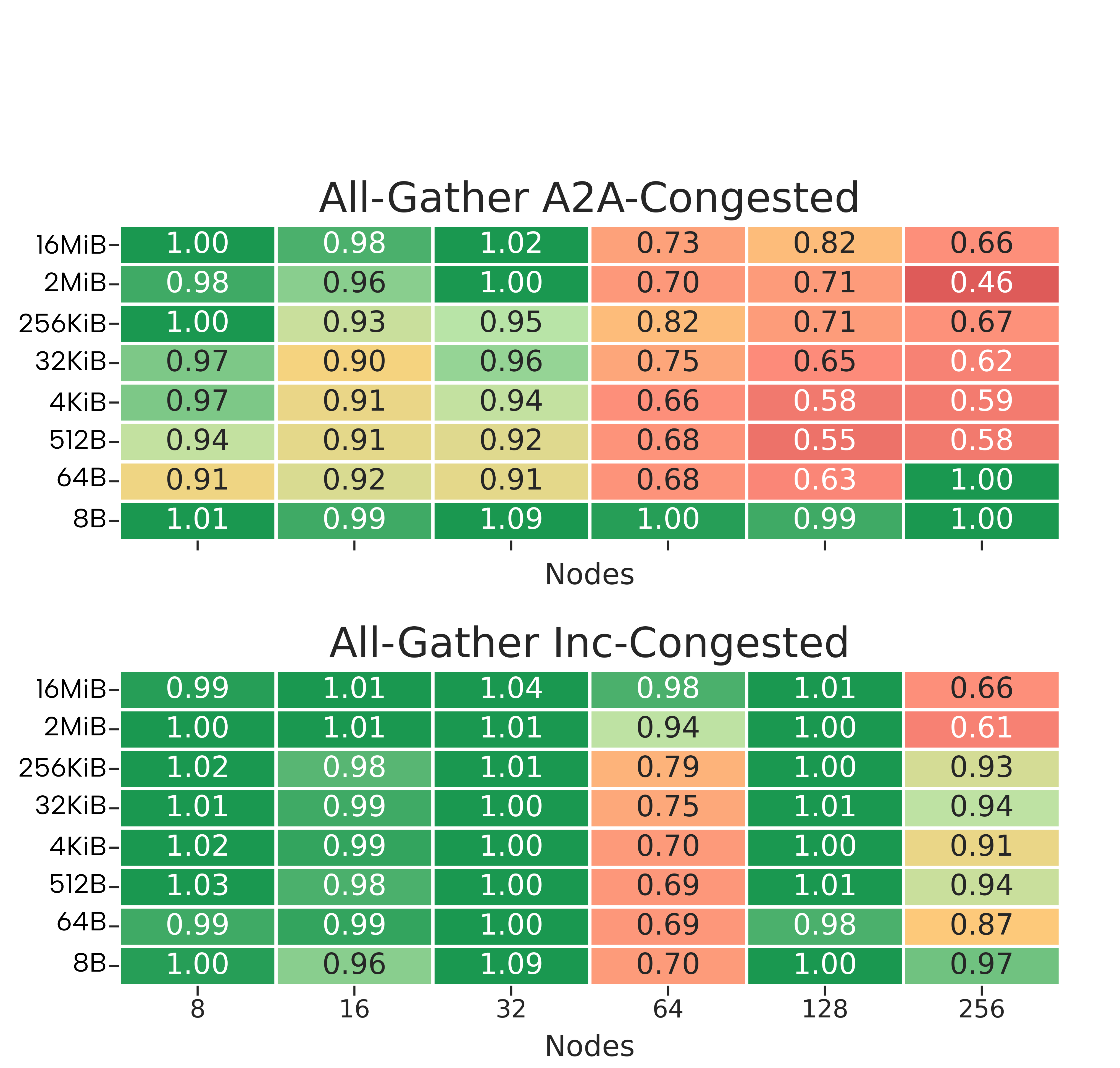}}%
  \hfill%
  \subfloat[Leonardo]{\includegraphics[width=0.32\textwidth]{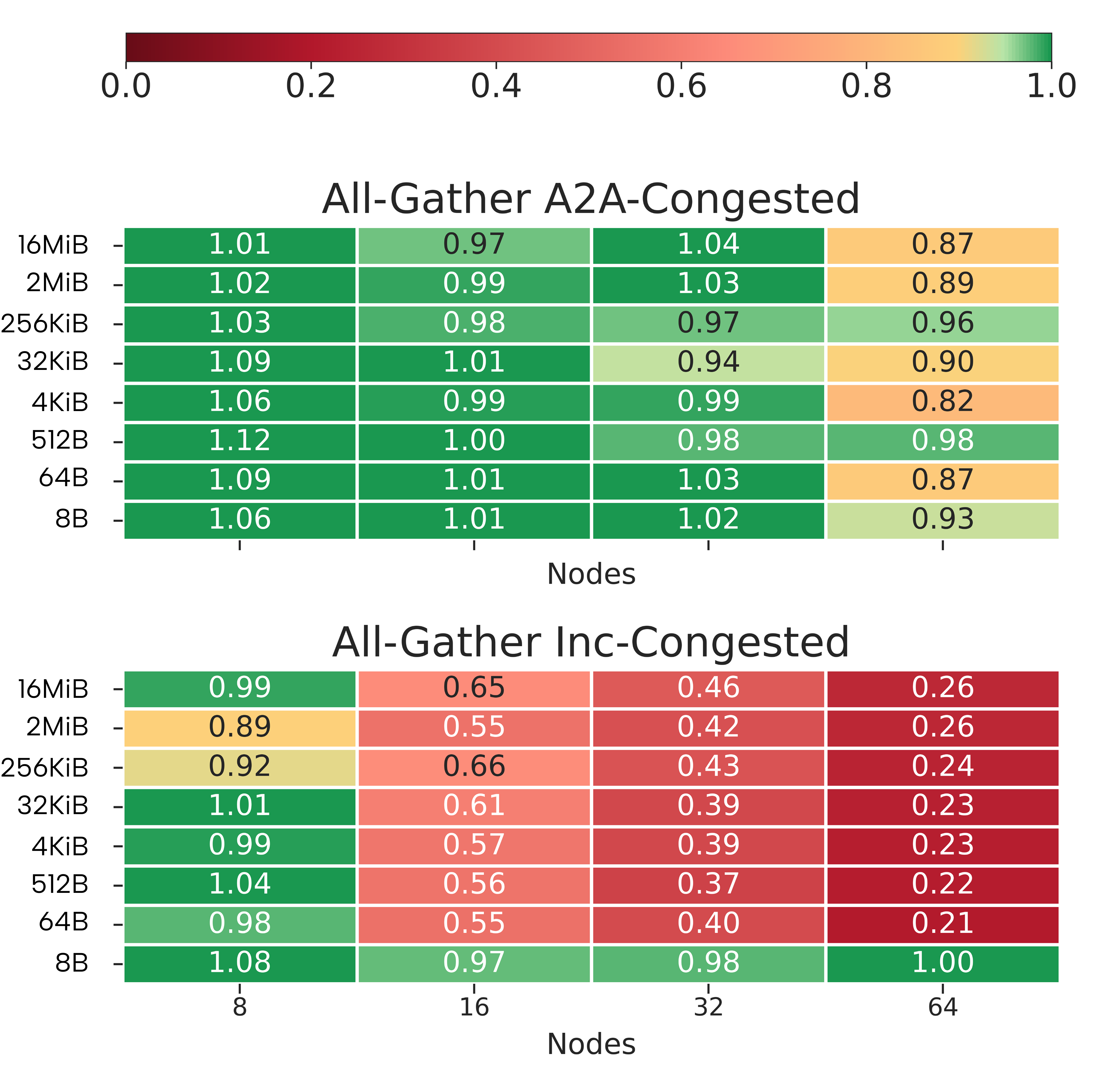}}%
  \hfill%
  \subfloat[LUMI]{\includegraphics[width=0.32\textwidth]{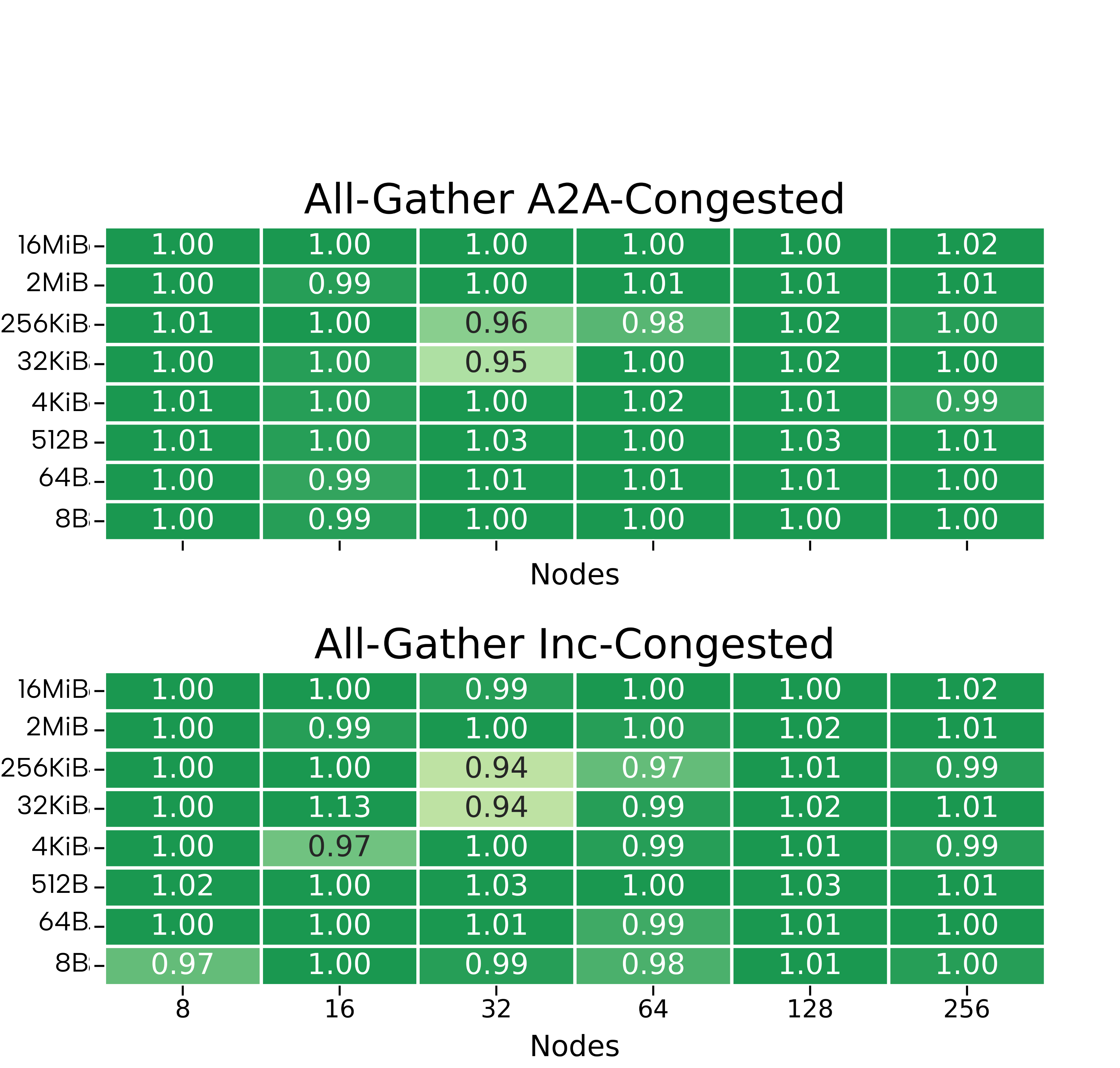}}
  \caption{Ratio between uncongested and congested runtimes on CRESCO8, Leonardo and LUMI, from 16 to 256 nodes, and vectors ranging from from 8 bytes to 16 MiB.}
  \label{fig:plot_scale}
\end{figure*}

\section{Steady Congestion at Large Scale}\label{sec:results:large}

After the small-scale study, we move to a more challenging regime by increasing the allocation up to 256 nodes, while keeping \textit{steady} and persistent congestion. In this section, we focus only on production supercomputers (Leonardo, CRESCO8, and LUMI), since the two Ethernet-based systems have no more than 10 nodes per partition. At 256 nodes, the experiment spans a non-negligible fraction of each machine: 7.39\% of Leonardo’s Booster partition, 8.60\% of LUMI’s GPU partition, and 33.68\% of the CRESCO8 CPU partition. 

These results are shown in Figure~\ref{fig:plot_scale}, where, for each system, we report two heatmaps: one showing the data from an \emph{AllGather} running together with an \emph{AlltoAll} aggressor, and the other showing the data from an \emph{AllGather} run with an \emph{Incast} aggressor. On each heatmap, we report on the x-axis the number of nodes (half used by the victim and half by the aggressor), and on the y-axis the size of the vector used as an input by the victim \emph{AllGather}. The value in each cell represents the ratio between the average uncongested and congested runtimes of the victim (i.e., the higher the better). 

It is worth noting that in some cases this might be slightly greater than $1$. I.e., due to temporary variability sometimes the congested execution, if not strongly affected by congestion, might be slightly faster (but still within the $10\%$) than the uncongested execution.

\subsection{CRESCO8}
\textbf{\emph{AlltoAll} Aggressor:} Figure~\ref{fig:plot_scale} (top left) highlights that CRESCO8 remains close to baseline under \textit{steady} \emph{\emph{AlltoAll}}-induced congestion up to 32 nodes, but a clear degradation emerges once allocations reach 64 nodes and beyond. For a few vector sizes, the value drops to as low as 0.45. This means that, when running with congestion, the application performance can be just the 45\% of the uncongested case, indicating that the fabric can no longer fully absorb the persistent congestion. The impact of congestion is even higher when using the all-to-all as a victim (not shown in the plot due to space constraints), with slowdowns up to $5\times$.

\textbf{\emph{Incast} Aggressor:} The effect is similar under \textit{steady} \emph{Incast} (Figure~\ref{fig:plot_scale}, bottom left), with performance dropping to 60\% of the uncongested case. However, the impact of congestion is larger when using the all-to-all as a victim (not shown in the plot due to space constraints), with slowdowns up to $25\times$.

\subsection{Leonardo}
\textbf{\emph{AlltoAll} Aggressor:} Figure~\ref{fig:plot_scale} (top center) shows a markedly different behavior on Leonardo. Under \textit{steady} \emph{AlltoAll} background traffic, performance remains close to (and in a few cases slightly above) the uncongested baseline across all tested scales: most entries stay around $\sim$0.95–1.05, with only minor localized slowdowns (e.g., $\approx$0.82–0.85) on $64$ nodes. This indicates that, for \textit{intermediate}-switch contention, Leonardo’s fabric provides sufficient path diversity and adaptive mechanisms to deflect traffic and preserve throughput, at least for the scales considered in our analysis.

\textbf{\emph{Incast} Aggressor:} The picture changes drastically with \textit{steady} \emph{Incast} (Figure~\ref{fig:plot_scale}, bottom center). While 8-node runs are still essentially unaffected, degradation appears already at 16 nodes and becomes severe at 32–64 nodes, where the ratio between uncongested and congested runtime for several vector sizes collapse to 0.2 (i.e., more than a $5\times$ slowdown). This is caused by \textit{edge}-localized congestion: the bottleneck is concentrated near the destination leaf switch, so alternative paths offer limited relief, and the performance is dominated by congestion control quality and buffering dynamics. 

\subsection{LUMI}
\textbf{\emph{AlltoAll} Aggressor:} LUMI exhibits the most robust behavior among the evaluated systems. As shown in Figure~\ref{fig:plot_scale} (top right), under \textit{steady} \emph{\emph{AlltoAll}} background traffic, the performance remains essentially unchanged across all tested scales, with performance in the congested case always within $5\%$ of the uncongested case. This indicates that \textit{intermediate}-switch contention is effectively managed by the fabric, and that load balancing preserves near-baseline performance even as the allocation grows.

\textbf{\emph{Incast} Aggressor:} More importantly, LUMI is also resilient to \textit{steady} \emph{Incast} (Figure~\ref{fig:plot_scale}, bottom right). Unlike Leonardo and CRESCO8, where \emph{Incast} triggers large collapses at moderate node counts, LUMI maintains ratios close to 1.0 across message sizes and node counts, with only minor fluctuations (generally within a few percent). 
This suggests that \textit{edge}-localized hot spots are handled effectively by the congestion control, which correctly identifies the victim flows and protect them from congestion induced by the \emph{Incast} aggressor. Overall, these results show that, within 8–256 nodes, LUMI sustains performance even in the challenging \textit{edge} congestion regime, providing stable throughput largely independent of both vector size and congestion pattern.

\begin{figure*}[!t]
  \centering
  \subfloat[CRESCO8]{\includegraphics[width=0.343\textwidth]{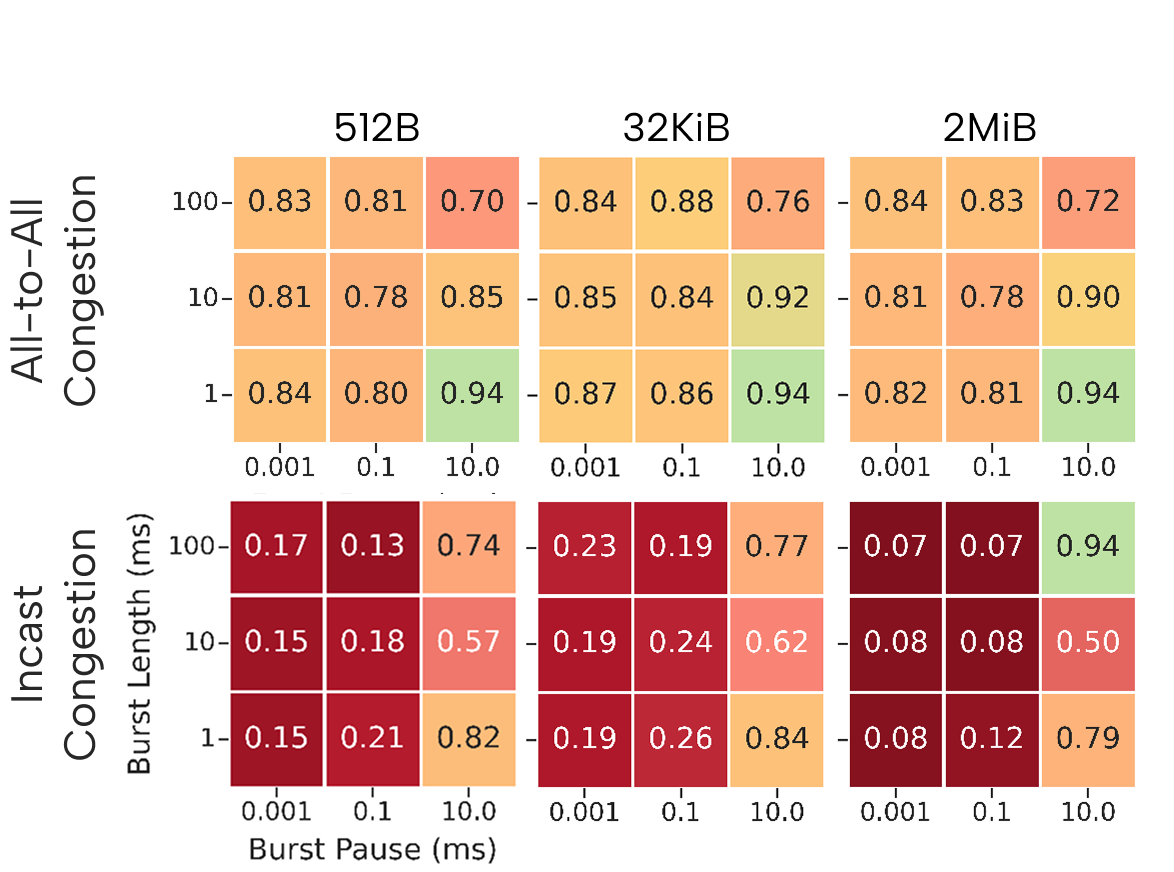}}%
  \hfill%
  \subfloat[Leonardo]{\includegraphics[width=0.31\textwidth]{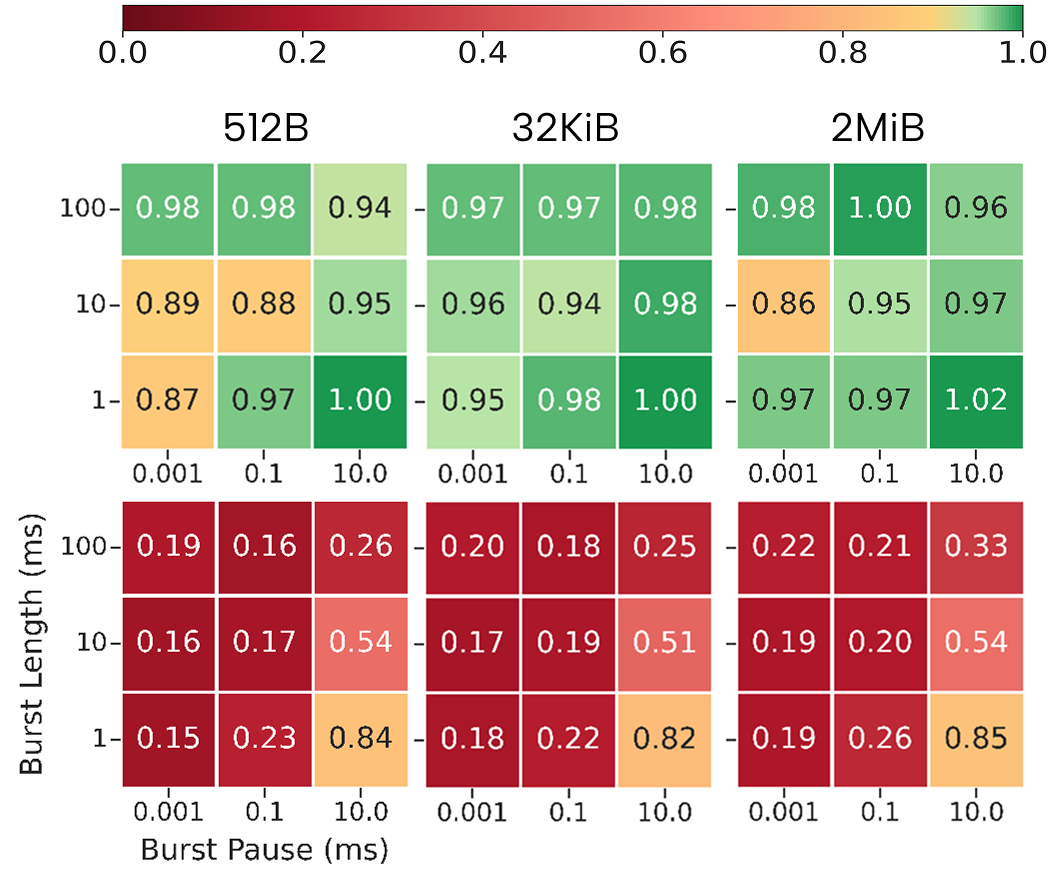}}%
  \hfill%
  \subfloat[LUMI]{\includegraphics[width=0.31\textwidth]{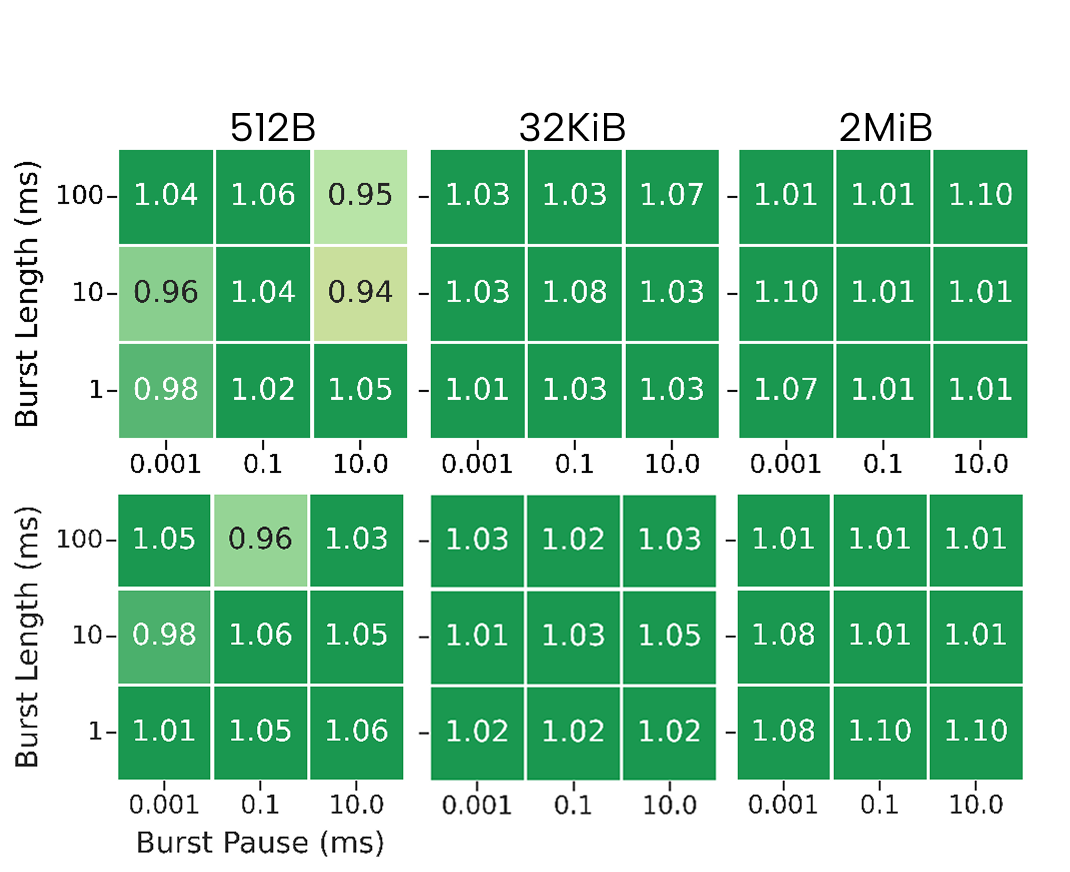}}
  \caption{Ratio between uncongested and congested runtime of 512 bytes, 32KiB and 2MiB \emph{AllGather} victim, on Leonardo, CRESCO8 and LUMI, for 64 nodes under \emph{AlltoAll} and \emph{Incast} aggressors.}
  \label{fig:64_compare}
\end{figure*}

\subsection{Summary}
Overall, the three supercomputers exhibit different behaviors, depending on their respective interconnect technologies and topologies. 
CRESCO8 deploys a blocking fat-tree, and while congestion does not significantly affect it below 32 nodes, performance degrades sharply as the allocation grows, in part also due to its tapered topology. This is particularly evident when using an \emph{AlltoAll} aggressor, especially since we are using a large fraction of the system (256 out of 760 nodes).

On the other hand, although Leonardo also relies on a tapered network (based on a Dragonfly+ topology), there are more opportunities to route traffic around hotspots since we are using a smaller fraction of the system. However, it suffers \emph{Incast} aggressors more than CRESCO8, highlighting issues related to congestion control. In contrast, LUMI combines the Cray Slingshot interconnect with a Dragonfly topology and shows the most stable behavior: in the presence of both \emph{AlltoAll} and \emph{Incast} congestion, the performance remains near baseline across scales, suggesting that its routing and congestion-management mechanisms handle both \textit{intermediate}- and \textit{edge}-localized \textit{steady} congestion more effectively.

\begin{observationbox}
\textbf{Observation 2.} 
Although CRESCO8 and Leonardo use similar network technology, they respond differently to different types of congestion. \textit{Alltoall} congestion has a higher impact on CRESCO8, whereas \textit{Incast} has a higher impact on Leonardo. This can be attributed in part to the smaller number of nodes in CRESCO8, in part to different network generation (older on Leonardo than on CRESCO8), and in part to adaptive routing and congestion control tuning or design differences. Lastly, it is worth noting that congestion (especially generated by \textit{Incast}) affects performance even at moderate allocation sizes (on the order of $\sim 2\%$ of a partition).
\end{observationbox}

\section{Bursty Congestion at Large Scale}\label{sec:results:bursty}

After characterizing \textit{steady} congestion, we move to a more realistic and challenging \textit{bursty} regime, where contention is injected intermittently with configurable burst duration and idle gaps, as described in Sec.~\ref{sec:cong:bursty}. \textit{Bursty} congestion is particularly challenging because it requires the fabric to operate in a dynamic environment: rather than converging once to a steady-state, the system must rapidly adapt to changes. We focus in this section on large-scale allocations, since the \textit{steady} congestion experiments showed that this is where congestion has the highest impact.

The results of this analysis on 64 nodes are shown in Fig.~\ref{fig:64_compare}, which displays multiple $3\times 3$ heatmaps, one for each aggressor type and victim vector size. Each of those heatmaps show the victim slowdowns (ratio between uncongested and congested runtimes), when varying the burst pause and the burst length. Moreover, we show similar results on 128 nodes on CRESCO8 in Fig.~\ref{fig:cresco_128}, and 256 nodes on LUMI in Fig.~\ref{fig:lumi_256}.

\subsection{CRESCO8}

\begin{figure*}[!t]
  \centering
  \includegraphics[width=1\textwidth]{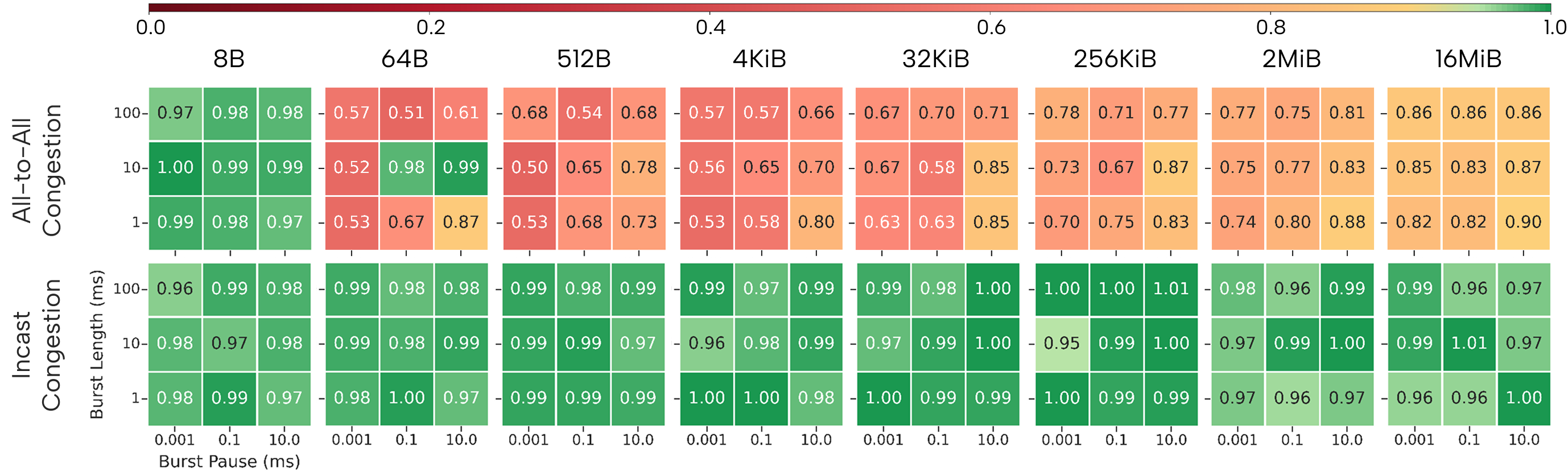}%
  \caption{Ratio between uncongested and congested runtime. 128 nodes \emph{AllGather} execution analysis on CRESCO8 under \emph{AlltoAll} and \emph{Incast} congestion ranging from 8 byte vectors to 16 MiB.}
  \hfill%
  \label{fig:cresco_128}
\end{figure*}

\textbf{\emph{AlltoAll} aggressor:} the performance degradation caused by \textit{bursty} \textit{intermediate} congestion is comparable to that of \textit{steady} congestion at the same node count. The application’s performance drops to $70\%$ of its uncongested baseline on 64 nodes (Fig.~\ref{fig:64_compare}) and 128 nodes (Fig.~\ref{fig:cresco_128}).

\textbf{\emph{Incast} aggressor:} Conversely, \emph{Incast} bursts are, in most cases, less tolerated on 64 nodes (Fig.~\ref{fig:64_compare}), with many configurations characterized by a ratio of approximately 0.08 (i.e., $12\times$ slower than the uncongested case). On 128 nodes (Fig.~\ref{fig:cresco_128}), performance is less affected, presumably because on a higher node count, the congestion tree originating from the Incast destination can spread over a higher number of switches, and thus the average queue length will be smaller.


\subsection{Leonardo}
\textbf{\emph{AlltoAll} Aggressor:}  Consistently with what observed in the \textit{steady} case, under \emph{AlltoAll} congestion, the system is largely resilient across most burst configurations: congested performance remains close to the uncongested baseline for small and medium message sizes, with only localized slowdowns appearing in the most aggressive regimes characterized by short gaps between bursts. 

\textbf{\emph{Incast} Aggressor:} The picture changes markedly for \emph{Incast} bursts. Here, performance degradation becomes widespread, especially when having short gaps between bursts, regardless of the burst length. Even in this case, a limited idle time prevents the system from recovering and dynamically adapting to congestion. On the other hand, long gaps between bursts, Leonardo experiences only a 15\% performance drop on short bursts, but a higher performance drops on longer bursts.


\begin{observationbox}
\textbf{Observation 3.} \textit{Bursty} congestion exposes the limits of reactive congestion handling: while modern HPC fabrics can often tolerate \textit{intermediate}-switch contention even under \textit{bursty} workloads, \textit{edge}-intensive congestion remains the dominant challenge. In addition, short idle gaps between bursts are especially harmful, leaving insufficient time for queues to drain and for endpoints and control mechanisms to react. Consequently, performance is bounded by edge buffering and endpoint rate control, for which additional path diversity provides limited mitigation.
\end{observationbox}

\subsection{LUMI}

\begin{figure*}[!t]
  \centering
  \includegraphics[width=1\textwidth]{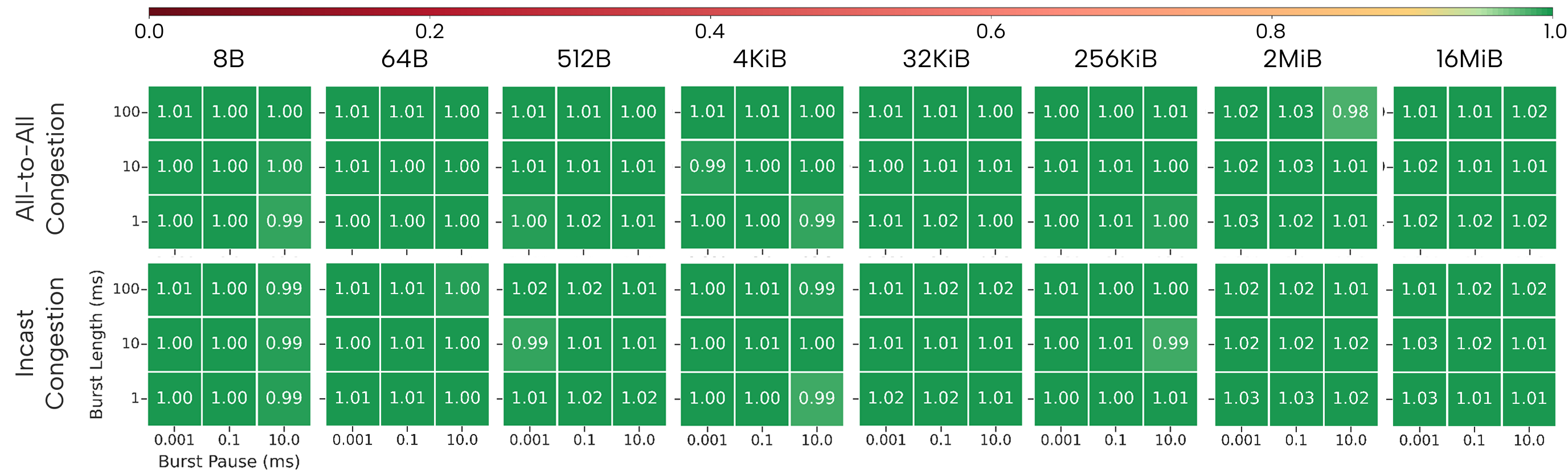}%
  \caption{Ratio between uncongested and congested runtime. 256 nodes \emph{AllGather} execution analysis on LUMI under \emph{AlltoAll} and \emph{Incast} congestion ranging from 8 byte vectors to 16 MiB.}
  \hfill%
  \label{fig:lumi_256}
\end{figure*}

\textbf{\emph{AlltoAll} Aggressor:} On both on 64 (Figure~\ref{fig:64_compare}) and 256 nodes (Figure \ref{fig:lumi_256}) LUMI exhibits the highest resilience to \textit{bursty} congestion among the evaluated systems. Across the explored burst configurations, throughput remains close to the uncongested baseline for almost all message sizes, indicating that transient congestion rarely escalates into sustained performance loss. Even for aggressive burst schedules (short gaps and longer bursts), the system rapidly absorbs and dissipates contention. 

\textbf{\emph{Incast} Aggressor:} Similar considerations apply for \emph{Incast} aggressors, regardless of scale, burst length, and burst idle gaps.

\begin{observationbox}
\textbf{Observation 4.} LUMI shows that Slingshot-based fabrics can effectively handle both \textit{intermediate} and \textit{edge} contention under dynamic, \textit{bursty} congestion, maintaining near-baseline throughput across a wide range of burst configurations.
\end{observationbox}

\subsection{Summary}
Results show that congestion resilience is not a single property of an interconnect, but an interaction between where contention occurs (\textit{intermediate} fabric vs. \textit{edge}-links), its pattern (\textit{steady} vs.\ \textit{bursty}), and the specific congestion-management mechanisms available at endpoints and in-network. High path diversity and adaptive routing (e.g., Dragonfly-like fabrics) are effective at absorbing \textit{intermediate} contention typical of \emph{AlltoAll}-like patterns, often preserving near-ideal throughput, unless a high fraction of the system is used (as we have seen on CRESCO8). 

However, \textit{edge}-intensive congestion remains the dominant challenge across platforms: it stress endpoint injection control and shallow leaf buffering, and can trigger sharp throughput collapse already at moderate allocation sizes (on the order of $\sim$2\% of a partition). \textit{Bursty} congestion further exposes the limits of reactive handling: performance becomes highly sensitive to the duty cycle, since short idle gaps do not provide enough time for buffers to drain and for rate control to converge. In this sense, the best-performing systems are not simply those with more connectivity, but those that combine path diversity with robust endpoint throttling and fast, stable congestion response.

While interconnect topologies may appear comparable, performance diverges based on the specific network generation and the congestion control and adaptive routing mechanisms. The design and fine-tuning of these algorithms are critical; they determine whether a system can maintain throughput or be affected by congestion under the unpredictable demands of multi-tenant workloads.

In production environments, these challenges are even more exacerbated by \textit{inter-job interference} from multiple co-running applications and the highly unpredictable, stochastic nature of real-world traffic. While our methodology employs \textit{periodic} bursts with fixed and recurring durations, representing a relatively 'favorable' or 'easy' scenario for network recovery, our side-by-side comparison reveals that even these regular patterns expose critical weak spots in modern \textit{adaptive routing} and \textit{congestion control} (CC) algorithms. The fact that these mechanisms struggle to stabilize the fabric under controlled conditions suggests that their efficacy in chaotic, production-grade environments remains a major concern.

\begin{observationbox}
\textbf{Observation 5.} 
Physical topology alone does not dictate how a system behaves under saturation; rather, congestion resilience is determined by the combined impact of the interconnect technology, generation, and topology. Ultimately, the system's performance depends on how these architectural features interact with the specific design and tuning of congestion control and adaptive routing algorithms. This multi-factored dependence explains why there is no single correlation between fabric layout and its ability to withstand network pressure.
\end{observationbox}

\section{Related Work}
The impact of network congestion and performance variability has been a significant focus of recent HPC research, ranging from empirical characterization to the development of system-level mitigation strategies.

\subsection{Characterization of Network Noise and Variability} Extensive research has focused on quantifying the impact of network noise using benchmarks centered on MPI collective operations~\cite{allreduce, htornnoise, Chunduri:2017:RVX:3126908.3126926, 1526010}. These studies establish a baseline for how variability affects communication performance in architectures like \textit{Cray Aries}. Recent comparative studies have also explored the performance gap between on-premise and cloud-based HPC environments, utilizing small-scale measurements and large-scale simulations to demonstrate how congestion can degrade collective performance as systems scale~\cite{cloudnoise}. Additionally, researchers have investigated the sensitivity of in-network collective offloading (where collective operations are processed within the switch) to background traffic and network contention~\cite{canary}.

\subsection{Job Placement and Adaptive Routing Strategies} A significant body of work explores how job allocation and placement strategies influence a network's sensitivity to congestion across various topologies, including Dragonfly~\cite{ibm:tm} and fat-trees~\cite{fattree:sc18, doi:10.1142/S0129626409000419}. These studies have identified that while randomized allocation can alleviate hotspots for communication-intensive workloads, contiguous placement is often superior for jobs with lower communication demands~\cite{bully, 8425264}.

Beyond placement, the role of adaptive routing in performance stability has been closely examined, particularly in low-diameter networks~\cite{sc2019}. Because routing decisions are typically made based on local switch visibility rather than a global network state, switches may select suboptimal, longer paths during periods of contention, which can result in significant performance drops depending on the application traffic pattern~\cite{sc2019, sc18}. Various telemetry tools have also been developed to aggregate hardware counters to help visualize and diagnose these network-wide congestion events~\cite{tool1, tool2, overtime}.

\subsection{Fabric-Specific Congestion Studies} Existing research often provides deep-dive analyses of specific hardware architectures. For instance, detailed studies have characterized the congestion management systems of the Slingshot and Aries fabrics~\cite{slingshot}, while other work has examined the impact of \emph{Incast} patterns on large-scale InfiniBand systems like \textit{Leonardo}, identifying how persistent traffic pressure reduces effective bandwidth~\cite{gpugpuinterconnect}.
\vspace{-.5em}
\subsection{Previous Works Limitations}
While the aforementioned studies were foundational, they often reflect the technological landscape of their time or focus on single architectures in isolation. Our work differs in several key aspects. On one side, whereas much of the existing literature relies on simulations or legacy technologies (e.g., \textit{Cray Aries}), this study provides a side-by-side experimental comparison of modern InfiniBand, Slingshot, and Ethernet ecosystems using actual production hardware. On the other side, by evaluating these fabrics under both steady-state and \textit{bursty} congestion, scenarios often overlooked in individual fabric studies, we provide a comprehensive view of interconnect resilience that is currently missing.

\section{Discussion and Conclusion}
Through a rigorous side-by-side evaluation of diverse interconnect architectures, this work identifies critical limitations in state-of-the-art congestion control and adaptive routing mechanisms. By employing a flexible methodology that subjects fabrics to both \textit{steady} and \textit{bursty} traffic profiles, we isolate weak spots where traditional mitigation strategies fail. Specifically, our results demonstrate that while path diversity effectively resolves \textit{intermediate} switch contention at small scales, \textit{edge} congestion remains a persistent bottleneck across several modern fabrics.
Furthermore, our analysis of \textit{bursty} traffic highlights that the network duty cycle—the relationship between burst intensity and recovery intervals—is the primary determinant of whether a fabric can maintain stability or falls into recurrent congestion cycles. These findings underscore that congestion management remains an open research challenge, suggesting that future system designs and procurement metrics should prioritize endpoint and leaf-link control over raw peak bandwidth.
More broadly, our analysis reveals that future fabrics must address the increasing communication bottleneck in AI and HPC workloads, which is exacerbated by congestion. Emerging standards such as Ultra Ethernet leverage standardized packet spraying to reduce the impact of congestion and mitigate tail latency. Meanwhile, other efforts are moving towards centralized adaptive load balancers specialized for LLM training. However, while centralized approaches can minimize congestion effects, their effectiveness at large scale or under highly dynamic traffic patterns remains limited.
As future work, we plan to extend the analysis to broader scales and incorporate application-driven traces to bridge controlled microbenchmarks with production workload behavior.

\section*{Acknowledgments}
We thank Matteo Marcelletti for the support with part of the code used for the benchmarks. This work is supported by the European Union’s Horizon Europe under grant 101175702 (NET4EXA), by Sapienza University Grants ADAGIO and D2QNeT (\textit{Bando per la ricerca di Ateneo} 2023 and 2024). We acknowledge ISCRA for awarding this project access to the LEONARDO supercomputer, owned by the EuroHPC Joint Undertaking, hosted by CINECA (Italy). We acknowledge the EuroHPC Joint Undertaking for awarding this project access to the EuroHPC supercomputer LUMI, hosted by CSC (Finland) and the LUMI consortium through a EuroHPC Regular Access call.
The CRESCO8 computing resources and related technical support used for this work were provided by the CRESCO/ENEAGRID HPC and its staff\cite{cresco}. CRESCO/ENEAGRID HPC infrastructure is funded by ENEA, the Italian National Agency for New Technologies, Energy and Sustainable Economic Development, and by Italian and European research programmes. Furthermore, we gratefully acknowledge the Goethe University of Frankfurt for hosting the HAICGU cluster.

\bibliographystyle{IEEEtran}
\bibliography{biblio}

@String{Computing = "Computing" }

@String{Computer = "{IEEE} Computer" }

@INPROCEEDINGS{cresco,
  author={F. {Iannone} and F. {Ambrosino} and G. {Bracco} and M. {De Rosa} and A. {Funel} and G. {Guarnieri} and S. {Migliori} and F. {Palombi} and G. {Ponti} and G. {Santomauro} and P. {Procacci}},
  booktitle={2019 International Conference on High Performance Computing   Simulation (HPCS)}, 
  title={{CRESCO ENEA HPC clusters: a working example of a multifabric GPFS Spectrum Scale layout}}, 
  year={2019},
  volume={},
  number={},
  pages={1051-1052},}

@misc{haicgu,
  title        = {HAICGU Cluster Documentation},
  author       = {{Open Edge and HPC Initiative (OEHI)}},
  howpublished = {\url{https://haicgu.github.io/index.html}},
  year         = {2022},
  note         = {Huawei AI and Computing at Goethe University (cluster based on Huawei HPC solution at Goethe University Frankfurt)}
}

@misc{TOP500,
  author       = {{TOP500 Project}},
  title        = {{TOP500: Ranking of the World’s 500 Fastest Supercomputers}},
  howpublished = {\url{https://top500.org/}},
  note         = {Accessed July 22, 2025},
  year         = {2025}
}

@misc{RowethCUG22Slingshot,
  author    = {Duncan Roweth},
  title     = {HPE Slingshot Launched into Network Space},
  howpublished = {Cray User Group (CUG) Proceedings},
  year      = {2022}
}

@inproceedings{RocherGonzalezCCGrid22,
  author    = {Jos{\'e} Rocher-Gonz{\'a}lez and Ernst Gunnar Gran and Sven-Arne Reinemo and Tor Skeie and Jes{\'u}s Escudero-Sahuquillo and Pedro Javier Garc{\'i}a and Francisco J. Quiles Flor},
  title     = {Adaptive Routing in InfiniBand Hardware},
  booktitle = {2022 22nd IEEE/ACM International Symposium on Cluster, Cloud and Internet Computing (CCGrid)},
  year      = {2022},
  pages     = {463--472},
  doi       = {10.1109/CCGrid54584.2022.00056}
}

@techreport{uec_spec,
  title       = {Ultra Ethernet Specification Version 1.0},
  institution = {Ultra Ethernet Consortium},
  year        = {2024},
  type        = {Technical Specification},
  note        = {Available from the Ultra Ethernet Consortium}
}

@inproceedings{gran2011ibcc,
  author    = {Ernst Gunnar Gran and Sven-Arne Reinemo},
  title     = {{InfiniBand Congestion Control: Modelling and Validation}},
  booktitle = {OMNeT++ 2011 (Workshop at SIMUTools)},
  year      = {2011},
  address   = {Barcelona, Spain},
  doi       = {10.4108/icst.simutools.2011.245509}
}

@inproceedings{alali2017ibcongestion,
  author    = {Fatma Alali and Fabrice Mizero and Malathi Veeraraghavan and John M. Dennis},
  title     = {{A Measurement Study of Congestion in an InfiniBand Network}},
  booktitle = {2017 Network Traffic Measurement and Analysis Conference (TMA)},
  year      = {2017},
  pages     = {1--9},
  doi       = {10.23919/TMA.2017.8002911}
}

@inproceedings{zhu2015dcqcn,
  author    = {Yibo Zhu and Haggai Eran and Daniel Firestone and Chuanxiong Guo and Marina Lipshteyn and Yehonatan Liron and Jitendra Padhye and Shachar Raindel and Mohamad Haj Yahia and Ming Zhang},
  title     = {{Congestion Control for Large-Scale RDMA Deployments}},
  booktitle = {Proceedings of the ACM SIGCOMM 2015 Conference (SIGCOMM '15)},
  year      = {2015},
  address   = {London, United Kingdom},
  pages     = {523--536},
  doi       = {10.1145/2785956.2787484}
}

@inproceedings{mittal2015timely,
  author    = {Radhika Mittal and Vinh The Lam and Nandita Dukkipati and Emily Blem and Hassan Wassel and Monia Ghobadi and Amin Vahdat and Yaogong Wang and David Wetherall and David Zats},
  title     = {{TIMELY: RTT-based Congestion Control for the Datacenter}},
  booktitle = {Proceedings of the ACM SIGCOMM 2015 Conference (SIGCOMM '15)},
  year      = {2015},
  address   = {London, United Kingdom},
  pages     = {537--550},
  doi       = {10.1145/2785956.2787510}
}

@inproceedings{li2019hpcc,
  author    = {Yuliang Li and Rui Miao and Hongqiang Harry Liu and Yan Zhuang and Fei Feng and Lingbo Tang and Zheng Cao and Ming Zhang and Frank Kelly and Mohammad Alizadeh and Minlan Yu},
  title     = {{HPCC: High Precision Congestion Control}},
  booktitle = {Proceedings of the ACM SIGCOMM 2019 Conference (SIGCOMM '19)},
  year      = {2019},
  address   = {Beijing, China},
  pages     = {44--58},
  doi       = {10.1145/3341302.3342085}
}

@inproceedings{10.1109/SCW63240.2024.0012,
author = {Pichetti, Lorenzo and De Sensi, Daniele and Sivalingam, Karthee and Nassyr, Stepan and Cesarini, Daniele and Turisini, Matteo and Pleiter, Dirk and Artigiani, Aldo and Vella, Flavio},
title = {Benchmarking Ethernet Interconnect for HPC/AI workloads},
year = {2025},
isbn = {9798350355543},
publisher = {IEEE Press},
url = {https://doi.org/10.1109/SCW63240.2024.00124},
doi = {10.1109/SCW63240.2024.00124},
booktitle = {Proceedings of the SC '24 Workshops of the International Conference on High Performance Computing, Network, Storage, and Analysis},
pages = {869–875},
numpages = {7},
location = {Atlanta, GA, USA},
series = {SC-W '24}
}

@misc{LUMI_website,
  title        = {{LUMI} Supercomputer},
  author       = {{LUMI Consortium}},
  howpublished = {\url{https://lumi-supercomputer.eu/}},
  year         = {2024},
  note         = {Accessed: 2025-01}
}

@article{CHIU19891,
title = {Analysis of the increase and decrease algorithms for congestion avoidance in computer networks},
journal = {Computer Networks and ISDN Systems},
volume = {17},
number = {1},
pages = {1-14},
year = {1989},
issn = {0169-7552},
doi = {https://doi.org/10.1016/0169-7552(89)90019-6},
url = {https://www.sciencedirect.com/science/article/pii/0169755289900196},
author = {Dah-Ming Chiu and Raj Jain}
}

@misc{ecmp,
        series =        {Request for Comments},
        number =        2992,
        howpublished =  {RFC 2992},
        publisher =     {RFC Editor},
        url =           {https://www.ietf.org/rfc/rfc2992.txt},
        author =        {C. Hopps},
        title =         {{Analysis of an Equal-Cost Multi-Path Algorithm}},
        year =          2009,
        month =         nov,
}

@inproceedings{Lu2018,
author = {Lu, Yuanwei and Chen, Guo and Li, Bojie and Tan, Kun and Xiong, Yongqiang and Cheng, Peng and Zhang, Jiansong and Chen, Enhong and Moscibroda, Thomas},
title = {Multi-path transport for RDMA in datacenters},
year = {2018},
isbn = {9781931971430},
publisher = {USENIX Association},
address = {USA},
booktitle = {Proceedings of the 15th USENIX Conference on Networked Systems Design and Implementation},
pages = {357–371},
numpages = {15},
location = {Renton, WA, USA},
series = {NSDI'18}
}

@misc{OpenMPI,
  title        = {{Open MPI: Open Source High Performance Computing}},
  howpublished = {\url{https://www.open-mpi.org/}},
  year         = {2025},
  note         = {Accessed: 2025-09-02},
  publisher    = {The Open MPI Project},
  address      = {Associated with Software in the Public Interest}
}

@article{Wang2025_NetworkLoadBalancing,
  title   = {Network Load Balancing Technologies for Intelligent Computing Centers},
  author  = {Weiguang Wang and Fei Chen and Peirui Cao and Liang Shan and Tao Wu and Huafeng Wen},
  journal = {Communications of Huawei Research},
  number  = {Issue 9},
  pages   = {13--22},
  year    = {2025},
  publisher = {Huawei Technologies Co., Ltd.},
  address   = {Shenzhen, China}
}

@INPROCEEDINGS{slingshot,
  author={De Sensi, Daniele and Di Girolamo, Salvatore and McMahon, Kim H. and Roweth, Duncan and Hoefler, Torsten},
  booktitle={SC20: International Conference for High Performance Computing, Networking, Storage and Analysis}, 
  title={An In-Depth Analysis of the Slingshot Interconnect}, 
  year={2020},
  volume={},
  number={},
  pages={1-14},
  doi={10.1109/SC41405.2020.00039}
}

@misc{OEHI_website,
  title        = {Open Edge and HPC Initiative},
  howpublished = {\url{https://www.open-edge-hpc-initiative.org/}},
  year         = {2025},
  note         = {Accessed: 2025-08-28},
}

@InProceedings{GPCNeT,
  author       = {Chunduri, Sudheer and Groves, Taylor and Mendygral, Peter and Austin, Brian and Balma, Jacob and Kandalla, Krishna and Kumaran, Kalyan and Lockwood, Glenn and Parker, Scott and Warren, Steven and Wichmann, Nathan and Wright, Nicholas J.},
  title        = {GPCNeT: Designing a benchmark suite for inducing and measuring contention in HPC networks},
  booktitle    = {Proceedings of the International Conference for High Performance Computing, Networking, Storage and Analysis (SC '19)},
  year         = {2019},
  publisher    = {Association for Computing Machinery},
  address      = {New York, NY, USA},
  articleno    = {42},
  numpages     = {33},
  doi          = {10.1145/3295500.3356215}
}

@inproceedings{plb,
author = {Qureshi, Mubashir Adnan and Cheng, Yuchung and Yin, Qianwen and Fu, Qiaobin and Kumar, Gautam and Moshref, Masoud and Yan, Junhua and Jacobson, Van and Wetherall, David and Kabbani, Abdul},
title = {PLB: congestion signals are simple and effective for network load balancing},
year = {2022},
isbn = {9781450394208},
publisher = {Association for Computing Machinery},
address = {New York, NY, USA},
url = {https://doi.org/10.1145/3544216.3544226},
doi = {10.1145/3544216.3544226},
booktitle = {Proceedings of the ACM SIGCOMM 2022 Conference},
pages = {207–218},
numpages = {12},
location = {Amsterdam, Netherlands},
series = {SIGCOMM '22}
}

@INPROCEEDINGS{Dixit2013,
  author={Dixit, Advait and Prakash, Pawan and Hu, Y. Charlie and Kompella, Ramana Rao},
  booktitle={2013 Proceedings IEEE INFOCOM}, 
  title={On the impact of packet spraying in data center networks}, 
  year={2013},
  volume={},
  number={},
  pages={2130-2138},
  doi={10.1109/INFCOM.2013.6567015}}

@inproceedings{flowbender,
author = {Kabbani, Abdul and Vamanan, Balajee and Hasan, Jahangir and Duchene, Fabien},
title = {FlowBender: Flow-level Adaptive Routing for Improved Latency and Throughput in Datacenter Networks},
year = {2014},
isbn = {9781450332798},
publisher = {Association for Computing Machinery},
address = {New York, NY, USA},
url = {https://doi.org/10.1145/2674005.2674985},
doi = {10.1145/2674005.2674985},
booktitle = {Proceedings of the 10th ACM International on Conference on Emerging Networking Experiments and Technologies},
pages = {149–160},
numpages = {12},
location = {Sydney, Australia},
series = {CoNEXT '14}
}

@misc{bonato2024smartt-anon,
      title={Under Submission}, 
      author={Details omitted for double-blind reviewing}
}

@misc{ultra,
      title={Ultra Ethernet}, 
      author={Ultra Ethernet Consortium},
      year={2024},
      note={https://ultraethernet.org/},
}

@inproceedings {flowlet,
author = {Erico Vanini and Rong Pan and Mohammad Alizadeh and Parvin Taheri and Tom Edsall},
title = {Let It Flow: Resilient Asymmetric Load Balancing with Flowlet Switching},
booktitle = {14th USENIX Symposium on Networked Systems Design and Implementation (NSDI 17)},
year = {2017},
isbn = {978-1-931971-37-9},
address = {Boston, MA},
pages = {407--420},
url = {https://www.usenix.org/conference/nsdi17/technical-sessions/presentation/vanini},
publisher = {USENIX Association},
month = mar
}

@inproceedings{flowcell,
author = {He, Keqiang and Rozner, Eric and Agarwal, Kanak and Felter, Wes and Carter, John and Akella, Aditya},
title = {Presto: Edge-Based Load Balancing for Fast Datacenter Networks},
year = {2015},
isbn = {9781450335423},
publisher = {Association for Computing Machinery},
address = {New York, NY, USA},
url = {https://doi.org/10.1145/2785956.2787507},
doi = {10.1145/2785956.2787507},
booktitle = {Proceedings of the 2015 ACM Conference on Special Interest Group on Data Communication},
pages = {465–478},
numpages = {14},
location = {London, United Kingdom},
series = {SIGCOMM '15}
}

@inproceedings{facebook,
author = {Gangidi, Adithya and Miao, Rui and Zheng, Shengbao and Bondu, Sai Jayesh and Goes, Guilherme and Morsy, Hany and Puri, Rohit and Riftadi, Mohammad and Shetty, Ashmitha Jeevaraj and Yang, Jingyi and Zhang, Shuqiang and Fernandez, Mikel Jimenez and Gandham, Shashidhar and Zeng, Hongyi},
title = {RDMA over Ethernet for Distributed Training at Meta Scale},
year = {2024},
isbn = {9798400706141},
publisher = {Association for Computing Machinery},
address = {New York, NY, USA},
url = {https://doi.org/10.1145/3651890.3672233},
doi = {10.1145/3651890.3672233},
booktitle = {Proceedings of the ACM SIGCOMM 2024 Conference},
pages = {57–70},
numpages = {14},
location = {Sydney, NSW, Australia},
series = {ACM SIGCOMM '24}
}

@inproceedings{hedera,
author = {Al-Fares, Mohammad and Radhakrishnan, Sivasankar and Raghavan, Barath and Huang, Nelson and Vahdat, Amin},
title = {Hedera: dynamic flow scheduling for data center networks},
year = {2010},
publisher = {USENIX Association},
address = {USA},
booktitle = {Proceedings of the 7th USENIX Conference on Networked Systems Design and Implementation},
pages = {19},
numpages = {1},
location = {San Jose, California},
series = {NSDI'10}
}

@inproceedings{conga,
author = {Alizadeh, Mohammad and Edsall, Tom and Dharmapurikar, Sarang and Vaidyanathan, Ramanan and Chu, Kevin and Fingerhut, Andy and Lam, Vinh The and Matus, Francis and Pan, Rong and Yadav, Navindra and Varghese, George},
title = {CONGA: distributed congestion-aware load balancing for datacenters},
year = {2014},
isbn = {9781450328364},
publisher = {Association for Computing Machinery},
address = {New York, NY, USA},
url = {https://doi.org/10.1145/2619239.2626316},
doi = {10.1145/2619239.2626316},
booktitle = {Proceedings of the 2014 ACM Conference on SIGCOMM},
pages = {503–514},
numpages = {12},
location = {Chicago, Illinois, USA},
series = {SIGCOMM '14}
}

@article{mptcp,
author = {Raiciu, Costin and Barre, Sebastien and Pluntke, Christopher and Greenhalgh, Adam and Wischik, Damon and Handley, Mark},
title = {Improving datacenter performance and robustness with multipath TCP},
year = {2011},
issue_date = {August 2011},
publisher = {Association for Computing Machinery},
address = {New York, NY, USA},
volume = {41},
number = {4},
issn = {0146-4833},
url = {https://doi.org/10.1145/2043164.2018467},
doi = {10.1145/2043164.2018467},
journal = {SIGCOMM Comput. Commun. Rev.},
month = {aug},
pages = {266–277},
numpages = {12}
}

@ARTICLE{ecnsu,
  author={Hoefler, Torsten and Roweth, Duncan and Underwood, Keith and Alverson, Robert and Griswold, Mark and Tabatabaee, Vahid and Kalkunte, Mohan and Anubolu, Surendra and Shen, Siyuan and McLaren, Moray and Kabbani, Abdul and Scott, Steve},
  journal={Computer}, 
  title={Data Center Ethernet and Remote Direct Memory Access: Issues at Hyperscale}, 
  year={2023},
  volume={56},
  number={7},
  pages={67-77},
  doi={10.1109/MC.2023.3261184}}

@misc{huaweiAIECN,
  title        = {AI ECN Threshold of Lossless Queues},
  author       = {{Huawei Support}},
  year         = {2024},
  howpublished = {\url{https://support.huawei.com/enterprise/en/doc/EDOC1100420118/7ade444e/ai-ecn-threshold-of-lossless-queues}},
  note         = {Accessed: 2025-12-20}
}

@ARTICLE{flowcut,
author={Bonato, Tommaso and De Sensi, Daniele and Di Girolamo, Salvatore and Bataineh, Abdulla and Hewson, David and Roweth, Duncan and Hoefler, Torsten},
journal={ IEEE Transactions on Networking },
title={{ Flowcut Switching: High-Performance Adaptive Routing With In-Order Delivery Guarantees }},
year={2025},
volume={},
number={01},
ISSN={2998-4157},
pages={1-14},
doi={10.1109/TON.2025.3636209},
url = {https://doi.ieeecomputersociety.org/10.1109/TON.2025.3636209},
publisher={IEEE Computer Society},
address={Los Alamitos, CA, USA},
month=dec}

@inproceedings{sc2019,
  author = {De Sensi, Daniele and Di Girolamo, Salvatore and Hoefler, Torsten},
  title = {Mitigating Network Noise on Dragonfly Networks Through Application-aware Routing},
  booktitle = {Proceedings of the International Conference for High Performance Computing, Networking, Storage and Analysis},
  series = {SC '19},
  year = {2019},
  isbn = {978-1-4503-6229-0},
  location = {Denver, Colorado},
  pages = {16:1--16:32},
  articleno = {16},
  numpages = {32},
  url = {http://doi.acm.org/10.1145/3295500.3356196},
  doi = {10.1145/3295500.3356196},
  acmid = {3356196},
  publisher = {ACM},
  address = {New York, NY, USA},
  openaccess = {https://dl.acm.org/authorize?N690645},
  dimensions = {true},
}

@INPROCEEDINGS{allreduce, 
author={T. Groves and Y. Gu and N. J. Wright}, 
booktitle={2017 IEEE International Conference on Cluster Computing (CLUSTER)}, 
title={Understanding Performance Variability on the Aries Dragonfly Network}, 
year={2017}, 
volume={}, 
number={}, 
pages={809-813}, 
doi={10.1109/CLUSTER.2017.76}, 
ISSN={2168-9253}, 
month={Sept},}

@INPROCEEDINGS{htornnoise, 
author={T. Hoefler and T. Schneider and A. Lumsdaine}, 
booktitle={2009 IEEE International Symposium on Parallel Distributed Processing}, 
title={The impact of network noise at large-scale communication performance}, 
year={2009}, 
volume={}, 
number={}, 
pages={1-8}, 
doi={10.1109/IPDPS.2009.5161095}, 
ISSN={1530-2075}, 
month={May},}

@inproceedings{Chunduri:2017:RVX:3126908.3126926,
 author = {Chunduri, Sudheer and Harms, Kevin and Parker, Scott and Morozov, Vitali and Oshin, Samuel and Cherukuri, Naveen and Kumaran, Kalyan},
 title = {Run-to-run Variability on Xeon Phi Based Cray XC Systems},
 booktitle = {Proceedings of the International Conference for High Performance Computing, Networking, Storage and Analysis},
 series = {SC '17},
 year = {2017},
 isbn = {978-1-4503-5114-0},
 location = {Denver, Colorado},
 pages = {52:1--52:13},
 articleno = {52},
 numpages = {13},
 url = {http://doi.acm.org/10.1145/3126908.3126926},
 doi = {10.1145/3126908.3126926},
 acmid = {3126926},
 publisher = {ACM},
 address = {New York, NY, USA},
}

@inproceedings{sc18,
 author = {Staci A. Smith and Clara E. Cromey and David K. Lowenthal and Jens Domke and Nikhil Jain and Jayaraman J. Thiagarajan and Abhinav Bhatele},
 title = {Mitigating Inter-Job Interference Using Adaptive
Flow-Aware Routing},
 booktitle = {Proceedings of the International Conference for High Performance Computing, Networking, Storage and Analysis},
 series = {SC '18},
 year = {2018},
}

@INPROCEEDINGS{1526010, 
author={D. {Skinner} and W. {Kramer}}, 
booktitle={IEEE International. 2005 Proceedings of the IEEE Workload Characterization Symposium, 2005.}, 
title={Understanding the causes of performance variability in HPC workloads}, 
year={2005}, 
volume={}, 
number={}, 
pages={137-149}, 
doi={10.1109/IISWC.2005.1526010}, 
ISSN={}, 
month={Oct},}

@article{canary,
  title = {Canary: Congestion-aware in-network allreduce using dynamic trees},
  journal = {Future Generation Computer Systems},
  volume = {152},
  pages = {70-82},
  year = {2024},
  issn = {0167-739X},
  doi = {https://doi.org/10.1016/j.future.2023.10.010},
  url = {https://www.sciencedirect.com/science/article/pii/S0167739X23003850},
  author = {{De Sensi}, Daniele and {Costa Molero}, Edgar and {Di Girolamo}, Salvatore and Vanbever, Laurent and Hoefler, Torsten},
  dimensions = {true},
}

@article{cloudnoise,
  author = {De Sensi, Daniele and De Matteis, Tiziano and Taranov, Konstantin and Di Girolamo, Salvatore and Rahn, Tobias and Hoefler, Torsten},
  title = {Noise in the Clouds: Influence of Network Performance Variability on Application Scalability},
  year = {2022},
  issue_date = {December 2022},
  publisher = {Association for Computing Machinery},
  address = {New York, NY, USA},
  volume = {6},
  number = {3},
  doi = {10.1145/3570609},
  journal = {Proc. ACM Meas. Anal. Comput. Syst.},
  month = dec,
  articleno = {49},
  numpages = {27},
  eprint = {2210.15315},
  dimensions = {true},
}

@inproceedings{ibm:tm,
 author = {Prisacari, Bogdan and Rodriguez, German and Heidelberger, Philip and Chen, Dong and Minkenberg, Cyriel and Hoefler, Torsten},
 title = {Efficient Task Placement and Routing of Nearest Neighbor Exchanges in Dragonfly Networks},
 booktitle = {Proceedings of the 23rd International Symposium on High-performance Parallel and Distributed Computing},
 series = {HPDC '14},
 year = {2014},
 isbn = {978-1-4503-2749-7},
 location = {Vancouver, BC, Canada},
 pages = {129--140},
 numpages = {12},
 url = {http://doi.acm.org/10.1145/2600212.2600225},
 doi = {10.1145/2600212.2600225},
 acmid = {2600225},
 publisher = {ACM},
 address = {New York, NY, USA},
}

@inproceedings{fattree:sc18,
 author = {Pollard, Samuel D. and Jain, Nikhil and Herbein, Stephen and Bhatele, Abhinav},
 title = {Evaluation of an Interference-free Node Allocation Policy on Fat-tree Clusters},
 booktitle = {Proceedings of the International Conference for High Performance Computing, Networking, Storage, and Analysis},
 series = {SC '18},
 year = {2018},
 location = {Dallas, Texas},
 pages = {26:1--26:13},
 articleno = {26},
 numpages = {13},
 url = {http://dl.acm.org/citation.cfm?id=3291656.3291691},
 acmid = {3291691},
 publisher = {IEEE Press},
 address = {Piscataway, NJ, USA},
}

@article{doi:10.1142/S0129626409000419,
author = {Bhatele, Abhinav and Kal\'{e}, Laxmikant V.},
title = {Quantifying Network Contention on Large Parallel Machines},
journal = {Parallel Processing Letters},
volume = {19},
number = {04},
pages = {553-572},
year = {2009},
doi = {10.1142/S0129626409000419},
URL = {https://doi.org/10.1142/S0129626409000419},
eprint = {https://doi.org/10.1142/S0129626409000419},
}

@INPROCEEDINGS{bully, 
author={X. Yang and J. Jenkins and M. Mubarak and R. B. Ross and Z. Lan}, 
booktitle={SC '16: Proceedings of the International Conference for High Performance Computing, Networking, Storage and Analysis}, 
title={Watch Out for the Bully! Job Interference Study on Dragonfly Network}, 
year={2016}, 
volume={}, 
number={}, 
pages={750-760}, 
doi={10.1109/SC.2016.63}, 
ISSN={2167-4337}, 
month={Nov},}

@INPROCEEDINGS{8425264, 
author={X. {Wang} and M. {Mubarak} and X. {Yang} and R. B. {Ross} and Z. {Lan}}, 
booktitle={2018 IEEE International Parallel and Distributed Processing Symposium (IPDPS)}, 
title={Trade-Off Study of Localizing Communication and Balancing Network Traffic on a Dragonfly System}, 
year={2018}, 
volume={}, 
number={}, 
pages={1113-1122}, 
doi={10.1109/IPDPS.2018.00120}, 
ISSN={1530-2075}, 
month={May},}

@article{tool1,
title = {Network Performance Counter Monitoring and Analysis on the Cray XC Platform.},
author = {Brandt, James M. and Froese, Edwin and Gentile, Ann C. and Kaplan, Larry and Allan, Benjamin A. and Walsh, Edward J},
abstractNote = {Abstract not provided.},
doi = {},
journal = {},
place = {United States},
year = {2016},
month = {5}
}

@INPROCEEDINGS{tool2, 
author={A. {Bhatele} and N. {Jain} and Y. {Livnat} and V. {Pascucci} and P. {Bremer}}, 
booktitle={2016 IEEE International Parallel and Distributed Processing Symposium (IPDPS)}, 
title={Analyzing Network Health and Congestion in Dragonfly-Based Supercomputers}, 
year={2016}, 
volume={}, 
number={}, 
pages={93-102}, 
doi={10.1109/IPDPS.2016.123}, 
ISSN={1530-2075}, 
month={May},}

@inproceedings{overtime,
 author = {Grant, Ryan E. and Pedretti, Kevin T. and Gentile, Ann},
 title = {Overtime: A Tool for Analyzing Performance Variation Due to Network Interference},
 booktitle = {Proceedings of the 3rd Workshop on Exascale MPI},
 series = {ExaMPI '15},
 year = {2015},
 isbn = {978-1-4503-3998-8},
 location = {Austin, Texas},
 pages = {4:1--4:10},
 articleno = {4},
 numpages = {10},
 url = {http://doi.acm.org/10.1145/2831129.2831133},
 doi = {10.1145/2831129.2831133},
 acmid = {2831133},
 publisher = {ACM},
 address = {New York, NY, USA},
}

@inproceedings{gpugpuinterconnect,
  author = {De Sensi, Daniele and Pichetti, Lorenzo and Vella, Flavio and De Matteis, Tiziano and Ren, Zebin and Fusco, Luigi and Turisini, Matteo and Cesarini, Daniele and Lust, Kurt and Trivedi, Animesh and Roweth, Duncan and Spiga, Filippo and Di Girolamo, Salvatore and Hoefler, Torsten},
  title = {Exploring GPU-to-GPU Communication: Insights into Supercomputer Interconnects},
  year = {2024},
  month = nov,
  booktitle = {Proceedings of the International Conference for High Performance Computing, Networking, Storage and Analysis (SC'24)},
  doi = {10.1109/SC41406.2024.00039},
  dimensions = {true},
}

@misc{turisini2023leonardopaneuropeanpreexascalesupercomputer,
      title={LEONARDO: A Pan-European Pre-Exascale Supercomputer for HPC and AI Applications}, 
      author={Matteo Turisini and Giorgio Amati and Mirko Cestari},
      year={2023},
      eprint={2307.16885},
      archivePrefix={arXiv},
      primaryClass={cs.DC},
      url={https://arxiv.org/abs/2307.16885}, 
}

\end{document}